\documentclass[11pt]{article} 
\usepackage{amsfonts} 
\usepackage{amsbsy} 
\usepackage{amssymb} 
\usepackage{graphicx} 
\begin{document} 
 
\title{BPS and non-BPS states in a \\ 
supersymmetric Landau-Ginzburg theory} 
\author{Alin Tirziu and Paul Fendley \\ 
\\ Department of Physics, University of Virginia \\ 
Charlottesville, VA 22904-4714} 
 
\maketitle 
 
\begin{abstract} 
We analyze the spectrum of the ${\cal N}=(2,2)$ supersymmetric 
Landau-Ginzburg theory in two dimensions with superpotential 
$W=X^{n+2}-\lambda X^{2}$. We find the full BPS spectrum of this 
theory by exploiting the direct connection between the UV and IR 
limits of the theory.  The computation utilizes results 
from the Picard-Lefschetz theory of singularities and its extension to 
boundary singularities. The additional fact that this theory is 
integrable requires that the BPS states do not close under 
scattering. This observation fixes the masses of non-BPS states as well. 
\end{abstract} 
 

 
\vskip .5in 
\section{Introduction} 
 
Supersymmetric field theories have many remarkable properties. In many 
cases, special quantities like the superpotential are holomorphic 
functions of the fields. The holomorphy makes it possible to 
obtain exact information about the theory. In many different 
types of theories, one can 
compute the exact spectrum of certain particles 
known as BPS states 
\cite{BPS}. Probably the most famous such computation is for 
four-dimensional ${\cal N}=2$ supersymmetric gauge theories, where 
Seiberg and Witten showed how to compute the BPS spectrum by 
exploiting the holomorphy of the superpotential and the fact that 
its singularity structure and monodromies can be inferred by 
perturbative arguments \cite{SW}. 
 
Before the work of \cite{SW}, a very close analog of this computation 
had already been done in ${\cal N}=(2,2)$ supersymmetric field 
theories in $1+1$ dimensions.  It was shown in \cite{CFIV} how one 
could derive a differential equation whose solution gives directly the 
exact BPS mass spectrum. The computation utilizes a technique called 
topological-antitopological fusion, introduced in \cite{CV1}. This 
work was extended further in \cite{CV2}, where it was shown how to 
extract the BPS spectrum without having to find the full solution of 
the differential equation. Instead, the BPS masses, a property of 
the infrared, are related
directly the charges of the fields 
in the Ramond sector at the ultraviolet conformal point.
 
For a variety of reasons, it seemed to us worthwhile to revisit these 
techniques. There has renewed interest in closely related two-dimensional 
theories with supersymmetry in the last few years (see e.g.\ 
\cite{Hori,Shifman,Malda}).  It is quite remarkable that exact 
non-perturbative information can be obtained in a strongly-interacting 
field theory, and in our opinion, the results of \cite{CV1,CFIV,CV2} 
have not yet been fully exploited.  Most of the examples studied in 
these earlier papers were fairly well understood prior to that work. In 
particular, many of them correspond to integrable field theories, for 
which there exist a number of standard techniques for deriving exact 
information. By using $S$-matrix techniques, the exact spectrum for 
many ${\cal N}=(2,2)$ theories had already been found in 
\cite{FMVW,FI1,FI2}.  Thus it would be useful to apply the results of 
\cite{CFIV,CV1,CV2} to cases where the answer is not known already. 
Moreover, even in the known cases, computing the spectrum is done 
indirectly -- one makes an ansatz and checks its consistency by the 
powerful constraints of integrability. The methods of 
\cite{CV2} allow a direct and exact computation of the BPS 
spectrum.

The key result exploited in this paper is quite simple to understand. 
The BPS states here form doublets under 
the supersymmetry, while all other states 
form a quartet. In ${\cal N}=(2,2)$ theories, there is a $U(1)$ 
fermion-number symmetry, denoted by $F$. The supercharges have fermion 
number $\pm 1$, so the particles in the BPS doublet have fermion 
numbers $(f,f-1)$ for some number $f$, while a quartet 
has charge $(f+1,f,f,f-1)$. 
The quantity introduced in \cite{CFIV} is 
a thermodynamic trace, 
$$C\propto \hbox{tr} \left[F(-1)^F e^{-\beta H}\right]$$ 
where $\beta$ the inverse temperature 
and $H$ the Hamiltonian on open space. The key property of 
$C$ is that only the BPS states appear in the 
one-particle contributions to this trace 
(the leading contributions at $\beta$ large). 
A BPS doublet of mass $m$ results in a one-particle contribution 
proportional to 
$$f(-1)^f + (f-1)(-1)^{f-1} = (-1)^f$$ 
to the trace. However, the quartet one-particle 
contribution vanishes, because $(f+1)-f-f+(f-1)=0$. As we will 
discuss, the fermion number $f$ and the mass $m$ of the BPS states are 
known from the superpotential.  Thus computing $C$ with a given set of 
boundary conditions gives the exact multiplicities of any BPS states 
obeying these boundary conditions. In fact, in \cite{CV2}, it was 
shown how to find the large-$m$ behavior of $C$ (and hence the 
multiplicities) without having to compute the whole thing.

The theories studied here correspond to superconformal minimal models 
perturbed by a relevant operator. In terms of a Landau-Ginzburg 
superfield $X$, the superpotential is $X^{n+2}-\lambda X^2$.  In this 
paper, we derive the exact BPS spectrum of these models, and use the 
integrability to learn about the non-BPS states.  In a sequel 
\cite{TF}, we will give their $S$-matrix. The series of models studied 
here is one of three sets of integrable ${\cal N}=(2,2)$ theories 
found by deforming the superconformal minimal model into a massive 
field theory \cite{FMVW,FLMW,MW}.  The other two deformations are well 
understood \cite{FI1,FI2}. The theory here is more complicated, and 
the $S$-matrix approach ran aground because of technical complications 
\cite{GS}.  The differential equations arising from 
topological-antitopological fusion were found in \cite{CV1}, but their 
general asymptotic behavior is not known, and so the BPS spectrum was 
not extracted.  The spectrum has been guessed at \cite{FLMW,LW}, but no 
authoritative results were known.  Another reason for doing our 
computation is that we will find some very interesting behavior. 
Due to the integrability, we can not only infer the 
masses of non-BPS states, but prove that they can be produced in a 
scattering event involving only the BPS states.

In section 2 we review the main characteristics of supersymmetric 
solitons in two dimensions, and discuss the theories with 
superpotential $W=X^{n+2}-\lambda X^{2}$.  In section 3 we review 
the work of \cite{CFIV,CV2}.  In particular we discuss the index 
$C$, and the relation between the BPS spectrum and the charges of 
the chiral fields. The proof and further use of this relation is 
directly related to singularity theory: the BPS spectrum is 
encoded in the monodromy matrix of the vanishing cycles. We 
recall, as needed, main results from the Picard-Lefschetz theory of 
singularities of differentiable maps \cite{picard1,picard2}.  The 
computation for $n$ odd in section 4 is a straightforward 
application of the above relation. The computation of BPS spectrum 
for $n$ even in section 5 requires additional mathematical 
results, from the singularity theory of functions on manifolds 
with boundary. In section 6 we prove the existence of non-BPS 
states in this theory, and present our conclusions.

\section{Supersymmetric solitons in 1+1 dimensions}

In this section we review  ${\cal N}=(2,2)$ supersymmetric solitons in 
1+1 dimensions, and introduce the theories studied in this 
paper. 

The two left supersymmetry operators
are denoted $Q^{+},Q^{-}$, and the two right are denoted 
$\overline{Q}^{+},\overline{Q}^{-}$. There is
a global $U(1)$ fermion-number symmetry called, generated by $F$.
We take space $\sigma$ to be the real line. 
When there are multiple vacua in a theory, there exist field 
configurations which are in one vacuum $a$ at $\sigma=-\infty$ and in 
another vacuum $b$ at the other end $\sigma=+\infty$.  These 
configurations are known as solitons, or in 1+1 dimensions, kinks.  
In the sector $ab$, the superalgebra is
\begin{eqnarray} 
\nonumber 
&\left\{ Q^{+},Q^{-}\right\} =H+P\quad & \left\{ \overline{Q}^{+}, 
\overline{Q}^{-}\right\} =H-P 
\\ 
&\left\{ Q^{+},\overline{Q}^{+}\right\} =\Delta \qquad\ & 
\left\{Q^{-},\overline{Q}^{-}\right\} =\overline{\Delta } 
\\ 
\nonumber 
&\left[ F,Q^{\pm }\right] =\pm Q^{\pm }\qquad\ & \left[ F, 
\overline{Q}^{\pm }\right] =\mp \overline{Q}^{\pm } 
\end{eqnarray} 
where $H$ and $P$ are the Hamiltonian and the momentum operators, and
all other anticommutators vanish.  The appearance of a central term
$\Delta_{ab}$ in superalgebras on non-compact spaces was noted long
ago \cite{olive}.  An important consequence of the central term is a
lower bound on the mass of any configuration in a given sector $ab$,
namely \cite{BPS,olive}
\begin{equation} 
m_{ab}\geqslant \left| \Delta _{ab}\right| .
\label{bound} 
\end{equation} 
The states whose mass saturates this bound are called BPS states. 
In an ${\cal N}=(2,2)$ theory in 1+1 dimensions, the BPS states 
have another very special property: they and only they form two-dimensional 
multiplets under the supersymmetry algebra. Other states form irreducible 
four-dimensional representations.

Many supersymmetric theories have an elegant Landau-Ginzburg 
description \cite{martinec,VW}.  An ${\cal N}=(2,2)$ Landau-Ginzburg 
theory is defined in terms of a set of $N$ chiral superfields 
$\phi^{i}$ each consisting of a complex boson and a Dirac fermion. The 
action is 
\[ 
S=\frac{1}{2}\int d^{2}zd^{4}\theta\ K\left(\phi^{i},\overline{\phi} 
^{i}\right)+\int d^{2}zd\theta^{+}d\theta^{-}\ W\left(\phi^{i}\right)+h.c. 
\] 
where $W$, the superpotential, is a holomorphic function and $K$ is a 
K\"{a}hler potential. 
The bosonic part of the LG action is given by: 
\[ 
S=\frac{1}{2}\int d^{2}z\left( g_{i \overline{j}}\partial _{\mu 
}X^{i}\partial _{\mu }\overline{X}^{j}+g^{i \overline{j}}\partial _{i}W 
\overline{\partial _{j}W}\right) 
\] 
where $X^{i}$ is the bosonic part of $\phi ^{i}$ and $g_{i\overline{j} 
}=\partial _{i} \overline{\partial} _{j}K$ is a positive-definite 
metric. Note the different kinds of derivatives: 
$\partial_\mu$ is the usual space-time derivative, whereas $\partial_i$ 
is a derivative with respect to the superfield $X^i$. 
The vacua $\left\{ 
a_{k}\right\}_{k=1,...,p} $ of the theory are obtained by 
minimizing the superpotential: 
\[ 
\frac{\partial W}{\partial X^{i}}\Bigg| _{a_{k}}=0 \quad\hbox{for all } 
i=1\dots N. 
\] 
The superpotential does not renormalize 
in a  ${\cal N}=(2,2)$ theory in two dimensions. Moreover, 
by dimensional analysis, it follows that the K\"ahler term is always 
irrelevant if $X$ has non-zero dimension. This means that critical 
points can be described uniquely by their superpotential.  One 
description of the superconformal minimal models is by a single 
superfield $X$ with $W=X^{n+2}$ \cite{martinec,VW}.  This 
is called the $A$ series; for $n$ even, one can orbifold by the $X\to 
-X$ symmetry and obtain the $D$ series. The $D$ superconformal 
series can be 
described by two superfields $X$ and $Y$ with superpotential 
$W=X^{k}-XY^2$.

We study time-independent solitons $X^{i}=X^{i}(\sigma )$, with 
$X^{i}(\sigma =-\infty )=a^{i}$ and $X^{i}(\sigma =\infty 
)=b^{i}$. The mass of such a configuration is 
\begin{eqnarray} 
\nonumber 
m_{ab}&=&\frac{1}{2}\int_{-\infty }^{\infty }d\sigma \left( g_{i\overline{j} 
}\partial _{\sigma }X^{i}\partial _{\sigma }\overline{X}^{j}+g^{i\overline{j} 
}\partial _{i}W\overline{\partial _{j}W}\right) 
\\ 
&=&\frac{1}{2}\int_{-\infty }^{\infty }d\sigma \left| \partial _{\sigma 
}X^{i}-\omega g^{i\overline{j}}\overline{\partial _{j}W}\right| 
^{2}+Re\left( \omega ^{*}\Delta W_{ab}\right) 
\label{solitonmass} 
\end{eqnarray} 
where $\Delta W_{ab}=W(b)-W(a)$, and $\omega$ is an arbitrary phase. Since 
the metric is positive-definite, the first term in (\ref{solitonmass}) 
is always greater than 
zero. Choosing 
$\omega =\Delta W/\left| \Delta W\right|$ 
yields the mass bound (\ref{bound}) 
with $\Delta_{ab}=\Delta W_{ab}$. 
The BPS solitons are those which saturate this bound: 
\begin{equation} 
m_{ab}=\left| \Delta W_{ab}\right|. 
\label{bpsmass} 
\end{equation} 
This computation is classical, 
but the crucial point is that the superpotential 
$W$ does not renormalize. Thus 
the exact non-perturbative mass of a BPS soliton is given by (\ref{bpsmass}). 
{}From (\ref{solitonmass}) we see that a BPS soliton must obey 
\begin{equation} 
\partial _{\sigma }X^{i}=\omega g^{i\overline{j}}\overline{\partial _{j}W} 
\label{solitoneqn} 
\end{equation} 
As a consequence, the superpotential for each soliton trajectory 
obeys 
$\partial _{\sigma }W=\partial _{i}W\partial _{\sigma }X^{i}=\omega \left| 
\partial W\right| ^{2}.$ 
This has a constant phase, so each BPS solution can be represented in 
$W$-plane as a straight line. Moreover, the mass is the length of the line.

In this paper we find the BPS spectrum 
of the Landau-Ginzburg theory with superpotential 
\begin{equation} 
W(X)=\frac{X^{n+2}}{n+2}-\lambda \frac{X^{2}}{2}. 
\label{superpot} 
\end{equation} 
For $\lambda=0$ this theory is the superconformal $A_{n+1}$ minimal series 
\cite{martinec,VW}. 
For $\lambda\neq 0$ the relevant perturbation $X^{2}$ 
 produces an integrable massive theory \cite 
{FLMW,MW}. The non-degenerate vacua are obtained by minimizing the 
superpotential $W(X)$, and in $X$-plane they are at the corners of an $n$ 
-polygon and the origin: 
\[ 
X_{0}=0;\qquad X_{k}=\lambda^{1/n}\exp \left( \frac{2\pi i}{n}k\right) 
\] 
where $k=1\dots n$. In $W$-plane the vacua are: 
\begin{equation} 
W_{0}=0,\quad W_{k}=-\frac{n}{2(n+2)}\lambda^{1+2/n} 
\exp \left( \frac{4\pi i}{n}k\right) 
\label{wvac} 
\end{equation} 
They are at the corners of an $n$-polygon and the origin for 
$n$ odd, and at the corners of an $n/2$-polygon and the origin for 
$n$ even. Because there is a $\mathbb{Z}_{n}$-symmetry acting transitively 
on the 
vacua in the $X$-plane, the soliton numbers between any two vacua $ 
X_{i},X_{j}$ except the origin depend only on $(i-j)$ mod $n$: 
\begin{equation} 
\mu _{ij}=\mu (j-i),\quad \quad  i\ne j\ne 0 
\label{solitonnums} 
\end{equation} 
with $\mu (i-n)=\mu (i)$.  The soliton numbers connecting the vacuum 
$X_{0}$ to $X_{k}$ are all equal due to the same 
$\mathbb{Z}_{n}$-symmetry. We denote this number by $\mu _{0k}\equiv 
\mu (0).$ There are thus $\left[ \frac{n}{2}\right] +1$ distinct 
soliton numbers to be computed, namely 
$\mu (0),\mu (1),....\mu \left( 
\left[ \frac{n}{2} \right] \right)$.

Another interesting feature of these BPS solitons is that 
their fermion number can be fractional, even though the supercharges 
have integer charge \cite{jack}. 
The fractional part of the fermion number 
$f_{ab}$ in the Landau-Ginzburg description is  \cite{FI2}: 
\begin{equation} 
e^{2\pi i f_{ab}}=\hbox{phase }\left[ \frac{\det H(b)}{\det H(a)}\right] 
\label{frac} 
\end{equation} 
where $H_{ij}=\partial_{i}\partial_{j}W$.  In the theory 
(\ref{superpot}), the states connecting the origin to another vacuum 
have fractional fermion number $\pm 1/2$. States connecting any two 
vacua except the origin have zero fractional fermion number, so these 
doublets consists of a boson and a fermion.

\section{Computing the BPS spectrum using Picard-Lefschetz theory}

The multiplicities and the masses of all BPS doublets in an 
${\cal N}=(2,2)$ supersymmetric theory can be computed directly 
and exactly. In this section we review the method, and set up the problem 
to be solved in the remainder of the paper. 
 
The Witten index \cite{witten} Tr$(-1)^{F}\exp(-\beta H)$ 
on closed space counts the vacua of the theory, and is independent 
of deformations of the theory. In \cite{CFIV}, an open-space 
generalization of this was introduced: 
\begin{equation} 
C_{ab}=\lim_{L\to \infty }\frac{i\beta 
}{2L}\,\hbox{Tr}_{ab}\left[ (-1)^{F}F e^{-\beta H}\right]. 
\label{index} 
\end{equation} 
where $L$ is the length of space.  The trace is over all states with 
boundary condition $a$ on the left and $b$ on the right.  Unlike the 
Witten index, $C_{ab}$ does depend on $\beta$.  It is an index in the 
sense that it does not depend on most deformations of theory; in a 
Landau-Ginzburg theory, it depends on the superpotential $W$, but not 
the K\"ahler term $K$.

The index $C_{ab}$ has simple expressions in both the ultraviolet 
and infrared limits of the theory.  The leading contribution in 
the infrared limit $\beta \gg 1$ are the one-particle states, 
since multi-particle states have in general higher energy. As 
discussed in the introduction, only the two-dimensional 
representation of the supersymmetry algebra thus contributes to 
the index in this limit.  In other words, 
\begin{equation} 
C_{ab}\Big|_{\beta \rightarrow \infty }=-\frac{i\beta}{2\pi }\mu _{ab}m_{ab} 
K_{1}(m_{ab}\beta ), 
\label{IR} 
\end{equation} 
where $\mu _{ab}$ is a real anti-symmetric matrix counting the 
number of BPS states between two vacua $a$ and 
$b$. Thus if $C_{ab}$ can be computed, 
one can read off the soliton numbers of all BPS states. 
This equation is a slight over-simplification, because there 
are special cases where two-particle states can also contribute to 
the index; these will be discussed below in section 5.2. 
The ultraviolet limit $\beta\to 0$ corresponds 
to studying the theory 
at the conformal point. In a massless theory, both left-moving 
and right-moving fermion numbers are conserved individually. 
In this limit, $C_{ab}$ gives the left (or 
right) charges of the Ramond vacua \cite{CFIV}: 
\[ 
C_{ab}\Big|_{\beta \rightarrow 0}=q_{ab}^{R} 
\] 
These charges can easily be determined from the Landau-Ginzburg 
superpotential \cite{martinec,VW}. For example, for the superconformal 
minimal models with superpotential $X^{n+2}$, the charges are 
\begin{equation} 
q_{m}^{R}=\frac{1}{2}\left( 1-\frac{2m}{n+2}\right) ,\quad\quad 
m=1\dots n+1 
\label{ramond} 
\end{equation} 
We will discuss in detail below how the $n+1$ Ramond charges 
are related to the different possible boundary conditions $(ab)$ on 
the solitons.

Using the technique of topological-antitopological fusion developed in 
\cite{CV1}, it was shown in \cite{CFIV} that one can derive
differential equations for the $C_{ab}$ in any ${\cal N}=(2,2)$ 
supersymmetric theory. In a Landau-Ginzburg theory, the superpotential 
gives these differential equations uniquely.  The Ramond charges serve as 
boundary conditions to determine the solution. One can extract 
the soliton numbers by solving these differential equations, but as 
discussed in detail in \cite{CFIV}, there is far more information in 
such a solution than just the soliton numbers. Moreover, the 
differential equations are non-linear, making the determination of 
asymptotic behavior (much less finding the entire solution)
a non-trivial task. To avoid this complication, it 
was shown in \cite{CV2} how to determine the $\beta\to\infty$ of the 
index directly, without having to find the full solution.  This has 
been proved for any ${\cal N}=(2,2)$ supersymmetric theory in two 
dimensions by using an integral formulation of topological 
anti-topological fusion equations. For Landau-Ginzburg theories the 
connection has been proved by using the Picard-Lefschetz theory of 
singularities of differentiable maps.

The problem is to compute the number of BPS soliton doublets 
between any pair of vacua $a$ and $b$. BPS solitons satisfy the 
equation (\ref{solitoneqn}). The idea is to study solutions of 
this equation near all the critical points, and continuously 
deform them away from the critical points. If two such solutions 
meet at some point in the $W$-plane along the line between two 
vacua, then there is a soliton. This must be a consistent solution 
because the equation (\ref{solitoneqn}) is of first order.

More precisely, 
we assume all critical points of $W$ are non-degenerate 
($\det(H_{ij})\neq 0$ 
at the critical points) and all critical values are distinct. 
By the Morse lemma there is local system of coordinates in the neighborhood of 
any critical point $a_i$ such that: 
\[ 
W(X^{1},...,X^{N})=W(a_i)+\sum_{i=1}^{N}(X^{i})^{2} 
\] 
Without loss of generality one can take vacuum $a$ at the origin in $X$ 
-plane with $W(a)=0$ and vacuum $b$ such that $W(b)$ is a real positive 
number. This means that $\omega=1$ for solutions of (\ref{solitonmass}) 
near $a$. 
Also, up to second order terms the metric 
can be chosen to be $g_{i\overline{j}}=2\delta _{i\overline{j}}$. 
The solution of equation (5) which as $\sigma =-\infty 
$ is at $a$ is: 
\[ 
X^{i}=\lambda ^{i}e^{\sigma }
\] 
where the $\lambda^i$ and hence the $X^{i}$ near $a$ must be real. 
The superpotential 
of any soliton solution is thus a real $N-1$-dimensional sphere with 
radius denoted by $\delta$. The $\lambda^{i}$ and hence 
the radius $\delta$ vanish right at the critical point $a$. 
This sphere comprises all possible 
soliton solutions near $a$, and we call it the \emph{wavefront} 
$\Delta_{a}$: 
\[ 
\Delta _{a}:\quad \sum_{i=1}^{N}(X^{i})^{2}=\delta 
\] 
Not for all $\lambda ^{i}$ there is a soliton solution: one must 
match the wavefront near $a$ to one near $b$. The 
wavefront near $b$ is defined analogously; all solutions of 
(\ref{solitoneqn}) near $b$ (here $\omega=-1$) form an 
$N-1$-dimensional sphere $\Delta _{b} $ which vanishes at $b$.

These definitions of $\Delta _{a},\Delta _{b}$ are exactly the 
definitions of vanishing cycles in singularity theory. 
We thus need to review some basics facts about 
the Picard-Lefschetz theory of 
singularities of differentiable maps \cite{picard2}. Assume a holomorphic 
function $W:\mathbb{C}^{N}\rightarrow \mathbb{C}$ has only non-degenerate 
critical points $a_{1},...,a_{p}$ whose critical values are all distinct $ 
W_{1},...,W_{p}$. If $t$ is a non-critical value in $W$-plane, the 
pre-image of it under $W$ is an $N-1$ complex dimensional space 
(non-singular level manifold) $V_{t}\equiv W^{-1}(t)$. One defines 
a system of non-self-intersecting paths $\left\{ \gamma 
_{i}\right\} _{i=1,...,p}$ to connect $t$ to each of the critical values $ 
W_{i}$: $\gamma _{i}:[0,1]\rightarrow \mathbb{C}$ with $\gamma 
_{i}(u=0)=W_{i},\gamma _{i}(u=1)=t$, $u\in [0,1]$. These paths do 
not cross any critical values. Also, they do not have to be 
cyclically ordered (weakly distinguished paths). 
For $u$ near zero one fixes in the level manifold 
$V_{\gamma(u)}$ an $N-1$ dimensional 
sphere: $S(u)=\sqrt{\gamma (u)-W_{i}}S^{N-1}$, where $ 
S^{N-1}$ is the standard $N-1$ dimensional sphere in a system of 
local coordinates near a critical point $a_{i}$: 
\[ 
S^{N-1}=\left\{ \left( X^{1},...,X^{N}\right) 
:\sum_{i=1}^{N}(X^{i})^{2}=1;\ 
\hbox{Im} X^{j}=0\hbox{ for all } j=1,..,N\right\} 
\] 
For $u=0$, the sphere $S(0)$ vanishes. Moving along the path 
$\gamma$ by homotopy lifting defines a family of $N-1$ dimensional 
cycles $S(u)\subset V_{\gamma(u)}$ diffeomorphic to spheres for 
all $u\in (0,1]$. Considering all critical points one obtains a 
set of $p$ vanishing cycles $\Delta _{i}$ represented by the 
spheres $S_{i}(1)$. These cycles are in the non-singular level set 
$V_{t}$ and they form a basis in the homology group 
$H_{N-1}(V_{t})$. Each homology 
class $\Delta _{i}$ is defined modulo orientation. The non-singular set $ 
V_{t}$ is homotopy equivalent to a bouquet of $p$ spheres each 
having a real dimension $N-1$. When $t$ tends to one of the 
critical values $W_{i}$ the corresponding cycle $\Delta _{i}$ 
vanishes. In $V_{t}$ the cycles can intersect. 
The intersection number of two cycles (denoted by $\circ$) 
is a bilinear 
operation defined on the homology group in accord 
with the preferred orientation on $V_{t}$. 
The orientation on $\mathbb{C} 
^{N} $ is defined by the positive orientation on $\mathbb{R}^{2N}$. For 
non-degenerate critical points with distinct critical values the intersection 
number takes values in the group of integers. In general the intersection 
number can take values in a subgroup of $\mathbb{C}$. 
 By comparing the two 
preceding paragraphs one sees that 
 the definition of cycles in singularity 
theory is indeed identical 
 to that of the solutions of (\ref{solitoneqn}) 
near a critical point.

The result of \cite{CV2} relates the number of soliton solutions 
between vacua $a$ and $b$ to the intersection number $\Delta_a\circ 
\Delta_b$ of the corresponding cycles.  We pick an arbitrary point $t$ 
on the line between $W(a)$ and $W(b)$, and define the paths $\gamma 
_{a},\gamma _{b}$ as straight lines connecting $t$ to $W_{a}$ and 
$W_{b}$ respectively.  The two wavefronts $\Delta _{a},\Delta _{b}$ 
are $N-1$ dimensional real cycles in $V_{t}$ for any $t$.  In 
$W^{-1}(t)$ the two vanishing cycles $\Delta _{a},\Delta _{b}$ 
intersect at a discrete number of points.  The fact that equation 
(\ref{solitoneqn}) is of first order assures that for each 
intersection point the flow with the vector field defined by 
(\ref{solitoneqn}) reaches the critical points $a$ and $b$. Therefore 
for each intersection point of vanishing cycles we get a soliton 
solution. However, the number of points at which $\Delta _{a}$ and $\Delta 
_{b}$ intersect is not necessarily $\Delta 
_{a}\circ \Delta _{b}$, because the latter counts each intersection 
point with $\pm 1$ depending on the orientations. Moreover, as 
suggested by the supersymmetric index $C_{ab}$, the counting of BPS 
solitons is naturally done with the insertion $(-1)^{F}F$.  Assume we 
have two different trajectories between vacua $a$ and $b$. Then, one 
has to compare the relative contribution of the two trajectories to 
$\Delta_{a}\circ \Delta_{b}$ and $(-1)^{F}F$. It was shown in 
\cite{CV2} that if the two trajectories correspond to the same sign 
for intersection between cycles they also have the same fermion number 
mod 2, and if they have opposite intersection number their fermion 
numbers differ by 1 mod 2. Also, including another vacuum $c$ the 
relative signs in the sectors $(ac),(cb)$ and $(ab)$ are correlated with 
$\Delta _{a}\circ \Delta _{c},\Delta _{c}\circ \Delta _{b}$ and 
$\Delta _{a}\circ \Delta _{b}$. Therefore the proper definition for 
the soliton number is \cite{CV2}: 
\begin{equation} 
\mu _{ab}=\sum_{(ab)} F(-1)^{F}=\Delta _{a}\circ \Delta_{b} 
\label{solitondefinition} 
\end{equation} 
where the sum is over all solitons in sector $(ab)$. 
This definition is valid for even number of fields. Because the intersection 
matrix of vanishing cycles for even number of fields is anti-symmetric the 
above definition is consistent with the physical requirement that the matrix 
$\mu_{ab}$ be anti-symmetric. For an odd number of fields the soliton numbers 
are 
\[ 
\mu _{ab}=\Delta _{a}\circ \Delta _{b}\quad \hbox{ for }\quad a>b,\quad \quad 
\mu_{ab}=-\Delta _{a}\circ \Delta _{b}\quad \hbox{ for }\quad a<b 
\] 
Since $|\mu_{ab}|$ counts the number of BPS doublets, 
$|\Delta _{a}\circ \Delta_b|$ does as well. 
 
The results of \cite{CV2} thus reduce the problem of finding the BPS 
spectrum to that of calculating the intersection numbers of the 
cycles. This can be done by studying the behavior of the cycles under 
deformation of the paths $\gamma_i$.  If the paths are continuously 
deformed the vanishing cycles do not change as long as the paths do 
not cross any critical value. But if during deformations a path 
crosses a critical value a change of basis appears. Let us consider 
two critical values $W_{i}$ and $W_{j}$ and two sets of paths $\gamma 
_{i},\gamma _{j}$ and $\gamma _{i}^{\prime },\gamma _{j}^{\prime }$ as 
in Figure 1. 
\begin{figure}[ht] 
\centerline{\includegraphics[scale=0.3]{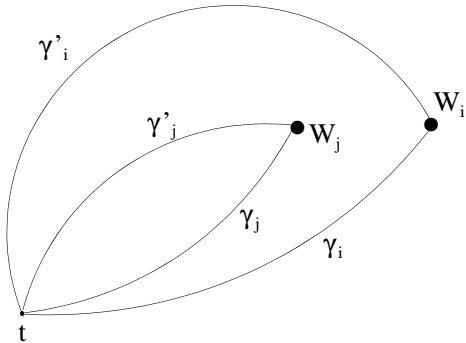}} \caption{Paths 
$\gamma_{i},\gamma_{j}$ change into 
$\gamma_{i}^{'},\gamma_{j}^{'}$} 
\end{figure} 
While $\gamma _{j}$ can be deformed into $\gamma _{j}^{\prime }$ 
without passing a critical value, the same is not true when deforming $ 
\gamma _{i}$ into $\gamma _{i}^{\prime }$. This change of paths can be 
viewed as an operation denoted by $\alpha _{j}(i)$ described as: $\gamma 
_{i}\rightarrow \gamma _{i}^{^{\prime \prime }}=\gamma _{i}\tau _{j}$ and $ 
\gamma _{j}\rightarrow \gamma _{j}^{^{\prime \prime }}=\gamma 
_{j}$, followed by a continuous change of paths: $\gamma 
_{i}^{^{\prime \prime }}\rightarrow \gamma _{i}^{^{\prime }}$ and 
$\gamma _{j}^{^{\prime \prime }}\rightarrow \gamma _{j}^{^{\prime 
}}$, where the \emph{simple loop} $\tau _{j}$ is defined as the 
loop going along the path $\gamma _{j}$ from the point $t$ to the 
point $W_{j}$, going around the point $W_{j}$ in the positive 
direction (anticlockwise) and returning along the path $\gamma 
_{j}$ to the point $t$ (see Figure 2). The inverse of operation 
$\alpha _{j}(i)$ is the operation $\beta _{j}(i)=\gamma _{i}\tau 
_{j}^{-1}$. 
\begin{figure}[ht] 
\centerline{\includegraphics[scale=0.3]{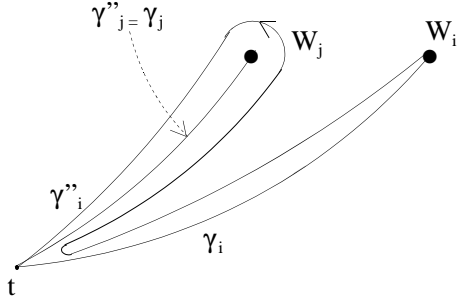}} \caption{Change of 
paths operation $\omega_{j}(i)$} 
\end{figure} 
A result of Picard-Lefschetz theory is that the basis of vanishing 
cycles changes under the operation $\alpha _{j}(i)$ as: 
\[ 
\Delta _{j}\rightarrow \Delta _{j}^{\prime }=\Delta _{j} 
\] 
\begin{equation} 
\Delta _{i}\rightarrow \Delta _{i}^{\prime }=\Delta 
_{i}+(-1)^{N(N+1)/2}(\Delta _{i}\circ \Delta _{j})\Delta _{j} 
\label{basischange} 
\end{equation} 
It is clear that the same change of basis of vanishing cycles 
obtained from operation $\alpha _{j}(i)$ is obtained by moving $t$ 
in clockwise direction as in Figure 3. 
\begin{figure}[ht] 
\centerline{\includegraphics[scale=0.3]{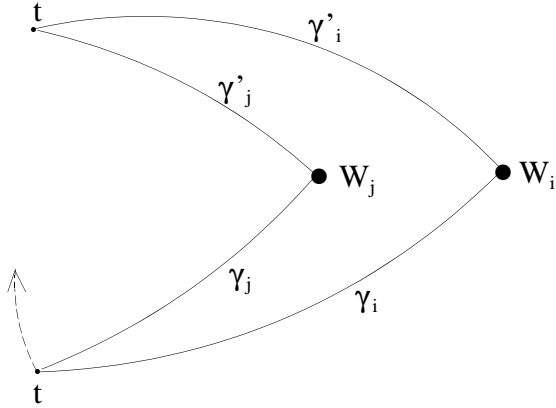}} \caption{The change 
of basis is given by $\alpha_{j}(i)$} 
\end{figure}

We can now define a monodromy matrix $M$, which describes how the 
basis of cycles changes as $t$ is deformed in a complete circle. 
Because of the basis-change formula (\ref{basischange}), the monodromy 
matrix depends on the intersection numbers.  However, its eigenvalues 
can be computed explicitly in terms of the Ramond charges, and so it 
will yield an equation for the soliton numbers.  To define $M$, we 
take the set of paths $\gamma _{i}$ to be straight lines (Figure 
4). Because the soliton numbers have a simpler definition for an even 
number of fields let us assume $N$ is even. 
If $t$ is along the 
straight line connecting vacua $W_{i}$ and $W_{j}$, one has $\mu 
_{ij}=\Delta _{i}\circ \Delta _{j}$. Suppose instead $t$ is taken to be very 
far from the critical values.  As $t$ is moved, $t$ will be collinear 
with pairs of vacua, and when this happens a change of basis occurs 
according to equation (\ref{basischange}). 
If $t$ goes completely around a large 
circle in clockwise direction with $\left| t\right| $ fixed, the 
resulting change of basis is called a monodromy. 
\begin{figure}[ht] 
\centerline{\includegraphics[scale=0.3]{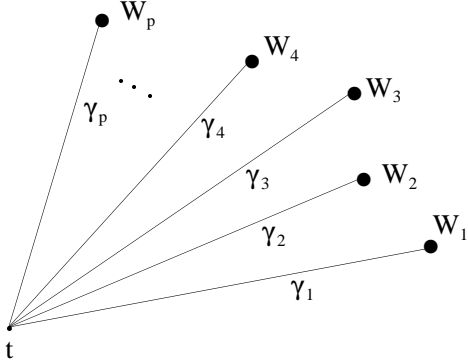}} \caption{The set of 
paths $\gamma_{i}$ connecting $t$ to the critical values $W_{i}$ 
are straight lines} 
\end{figure} 
Let $A_{ij}$ be the matrix whose 
only non-zero entry is the $(i,j)$ entry, which is equal to $\mu 
_{ij}=\Delta _{i}\circ \Delta _{j}$.
The monodromy matrix $M$ is then an 
ordered product of change-of-basis matrices 
$M_{ij}=I+(-1)^{N(N+1)/2}A_{ij}$. As $t$ goes halfway around the 
circle it will be collinear with all possible pairs of critical 
values; there are $p(p-1)/2$ 
such pairs. We denote $S$ to be the ordered product of matrices 
$M_{ij}$ as $t$ goes half the way around the circle by: 
\[ 
S=\prod_{pairs\mathtt{\ }(i,j)}^{\rightarrow }M_{ij} 
\] 
Since $M_{ji}=M_{ij}^{-t}$, as $t$ goes around the 
full circle the monodromy matrix is: 
\[ 
M=SS^{-T} .
\] 
 
The monodromy matrix contains information about the BPS spectrum, 
namely the soliton numbers between all pairs of vacua.  The 
crucial point is that it can be computed in a completely different 
way \cite{CV2}. It is known from singularity theory that the 
monodromy matrix is independent of finite deformations of the 
vacua.  Moreover, one can compute the eigenvalues of the monodromy 
matrix directly in the superconformal limit $\lambda\to 0$, where 
all the $W_i$ are the same. They can be expressed in terms of 
$U(1)$ charges of the Ramond vacua $q_{m}^{R}$.  For an even 
number of fields, one has \cite{CV2} 
\begin{equation} 
\hbox{eigenvals}(M)=\exp (2\pi iq_{m}^{R}). 
\label{evM} 
\end{equation} 
This is the key equation: the left-hand-side can be expressed solely 
in terms of the BPS soliton numbers, while the right hand side gives 
the spectrum of charges of chiral fields of the superconformal theory. 
This equation is quite profound: 
it directly relates a property of the infrared (the soliton spectrum) 
to that of the ultraviolet (the Ramond charges).

The assumption that the number of fields is even is not 
a restriction, because one can always add a non-interacting 
field $Y^{2}$ to the 
superpotential. Thus, for convenience in defining the soliton numbers in 
terms of intersection matrix of vanishing cycles, we consider the 
modified superpotential: 
\[ 
W(X,Y)=\frac{X^{n+2}}{n+2}-\lambda \frac{X^{2}}{2}+Y^{2} .
\] 
This modified superpotential and the original one have the same 
$X$-plane and $W$-plane vacua geometry. The addition of the field 
$Y^{2}$ only changes signs in the intersection matrix, 
and the actual soliton numbers are $|\mu_{ab}|$. Since for 
functions of an even number of variables the self-intersection matrix 
of vanishing cycles is zero, the physical requirement $ 
\mu_{aa}=0$ follows directly. The exponentials $\exp (2\pi iq_{m}^{R})$ 
of the Ramond charges (\ref{ramond}) 
for this theory are the zeroes of the polynomial $P(z)$ 
\begin{equation} 
P(z)=z^{n+1}-z^{n}+z^{n-1}-\dots\pm 1. 
\end{equation} 
with the last sign $+$ for $n$ odd and $-$ for $n$ even. 
Thus the relation (\ref{evM}) for the soliton numbers 
can be rewritten as: 
\begin{equation} 
\det(M-z)=P(z) .
\label{detM} 
\end{equation} 
 
To complete the computation we need to compute the monodromy matrix 
$M$. We do this in the next two sections. We treat separately the 
cases $n$ odd and $n$ even, because the geometry of the vacua in 
$W$-plane as well as the definitions 
of the soliton numbers differ slightly in the two cases.

\section{The BPS spectrum for $n$ odd} 
 
In this section we compute the soliton numbers 
(\ref{solitonnums}) $\mu(i)$ in the case $n$ odd. 
When $n$ is odd, the critical values (vacua) (\ref{wvac}) in $W$-plane are 
all distinct.  The $ \mathbb{Z}_{n}$-symmetry $X\to \exp(2\pi ik/n) X$ in 
$X$-plane results in a $\mathbb{Z}_{n}$-symmetry in the $W$-plane 
which takes $W\to \exp(4\pi ik/n) W$. 
The soliton numbers are defined by (\ref{solitondefinition}) in terms of 
the intersection matrix of vanishing cycles. The modified superpotential 
contains an even number of fields. Therefore the soliton numbers satisfy: 
\[ 
 \mu (-i)=-\mu (i),\quad\quad 
i=1,...,\left[ n/2\right]; 
 \] 
\[ 
\mu _{0i}=-\mu _{i0}=\mu(0),\quad\quad i=1,\dots,n. 
\]

To compute the monodromy matrix $M$, we take a non-vacuum point $t$ 
in $W$-plane far from the vacua just below the real axis with 
$Re(t)<0$, and let it move in a clockwise direction around the 
critical values with $\mid t\mid =$ fixed (see figures 5 and 6). We 
consider all lines connecting the vacua which intersect, 
in order, the half circle obtained by rotating $t$ by $\pi$.
There are $(n+1)/2$ lines passing through each of the $n$ points 
(corners of the $n$-polygon) in $W$-plane, each line corresponding 
to one of $\mu (0),$ $\mu (1),......\mu (\left[ \frac{n}{2}\right] )$. 
 
\begin{figure}[ht] 
\centerline{\includegraphics[scale=0.4]{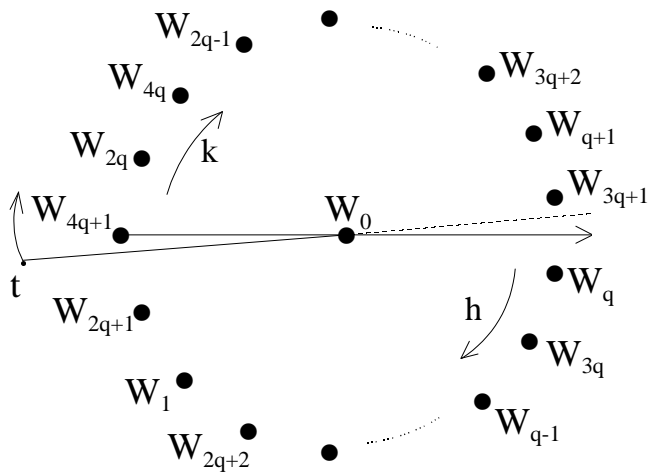}} \caption{Vacua 
geometry in $W$-plane for $n=4q+1$} 
\end{figure} 
\begin{figure}[ht] 
\centerline{\includegraphics[scale=0.4]{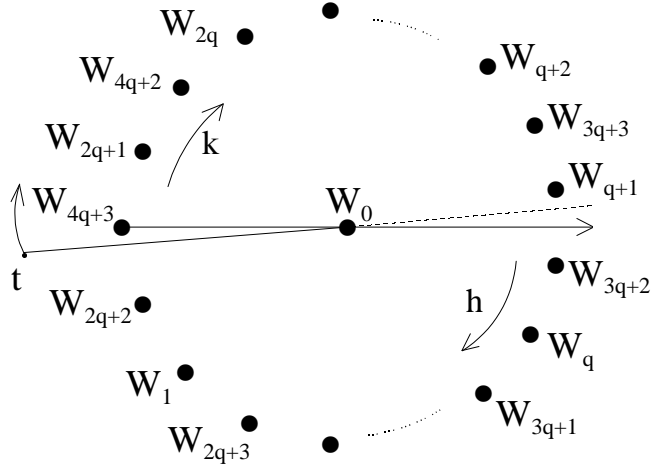}} \caption{Vacua 
geometry in $W$-plane for $n=4q+3$} 
\end{figure} 
 
The matrix $S$ consists of a product of $n(n+1)/2$ change of basis 
matrices $M_{ij}=I-A_{ij}$. However, we can find a commuting set 
of $(n+1)/2$ matrices $A_{ij}$ which contain each different 
soliton number $\mu(i)$ once.  This suggests that one may be able 
to simplify the matrix $S$, and thus the monodromy matrix 
$M=SS^{-T}$, so that (\ref{detM}) can be solved for soliton 
numbers. Let us look at vacua in $W$-plane, and consider a line 
passing through any two vacua except the origin. There is a set of 
$(n-1)/2$ such lines parallel to each other. If the point $t$ is 
far enough from the vacua, it crosses all these parallel lines 
sequentially as it is rotated. Moreover, the product of $A$ 
matrices corresponding to any two of these lines must vanish, 
since the parallel lines join different critical points. We add to 
this set the line through the origin and another critical point 
which appears (as $t$ is rotated) just before these parallel 
lines. The $A$ matrix for this line also gives zero when 
multiplied with the others in the set. This set of $(n+1)/2$ lines 
therefore contributes to the matrix $S$ as $I-\mu ^{<s>}$, where 
$\mu ^{<s>}$ is the sum of all $A$ matrices associated with this 
set. There are $n$ such sets of lines, so the matrix $S$ can be 
written as: 
\begin{equation} 
S=(I-\mu ^{<1>})(I-\mu ^{<2>}).........(I-\mu ^{<n>}) 
\end{equation} 
To give an  example, for $n=7$, the vacua form a pentagon, 
as displayed in figure 7 below. 
There are three $\mu's$ to be computed: $\mu(0),\mu(1),\mu(2)$. 
The matrix $S$ contains five sets each consisting of three 
matrices: 
\begin{eqnarray*} 
S=(M_{05}M_{15}M_{42})(M_{10}M_{35}M_{12}) 
(M_{02}M_{32}M_{14})(M_{30}M_{52}M_{34 
})(M_{04}M_{54}M_{31}) 
\end{eqnarray*} 
where $M_{ij}=I-A_{ij}$. The product of any two matrices $A$ in 
each set gives zero. Then: 
\begin{eqnarray*} 
S&=&(I-A_{05}-A_{15}-A_{42})(I-A_{10}-A_{35}-A_{12})(I-A_{02}-A_{32}-A_{14}) \\ 
& &(I-A_{30}-A_{52}-A_{34})(I-A_{04}-A_{54}-A_{31}) 
\end{eqnarray*} 
Here 
\[\mu^{<1>}=A_{05}+A_{15}+A_{42},...,\mu^{<5>}=A_{04}+A_{54}+A_{31}\]

To simplify $S$ further, we 
note that in each of these sets $\mu^{(s)}$, 
each different soliton number $\mu(i)$ 
($i =0, 1\dots (n-1)/2 $) appears exactly once. 
We can connect any matrix $A$ belonging to $\mu ^{<s+1>}$ 
to one of $\mu ^{<s>}$ in the same way, which then provides a 
connection between $\mu ^{<s+1>}$ and $\ \mu ^{<s>}$. 
Precisely, we will show that 
the monodromy matrix $M$ can be reduced to 
\begin{eqnarray} 
\label{s1} 
M &=&\left[ (I-\mu ^{<1>})R^{q}(I+\mu ^{<1>})^{T}R^{q}\right] ^{n} 
\qquad\quad \qquad \hbox{ for } n=4q+1 \\ 
\label{s2} 
M&=&\left[ (I-\mu 
^{<1>})R^{-q-1}(I+\mu ^{<1>})^{T}R^{-q-1}\right] ^{n} \mathtt{\ } 
\qquad\hbox{ for } n=4q+3 
\end{eqnarray} 
where $R$ is the $(n+1)\times (n+1)$ 
matrix: 
\[ 
R=\left( 
\begin{array}{lllllll} 
1 & 0 & 0 & 0 & ... & 0 & 0 \\ 
0 & 0 & 0 & 0 & ... & 0 & 1 \\ 
0 & 1 & 0 & 0 & ... & 0 & 0 \\ 
0 & 0 & 1 & 0 & ... & 0 & 0 \\ 
.. & .. & .. & .. & ... & .. & .. \\ 
.. & .. & .. & .. & .. & .. & .. \\ 
0 & 0 & 0 & 0 & 0 & 1 & 0 
\end{array} 
\right) 
\] 
$R$ satisfies: $R^{-n}=R^{n}=I$ and $R^{T}=R^{-1}$.

To derive (\ref{s1}) and (\ref{s2}), we use the notation 
$\langle r,s \rangle$ for the ray starting at $r$ and passing 
through $s$. There are $\frac{n-1}{2}$ points for which $\hbox{Im}(W_{k})>0$ and 
one point, $k=n$, for which $\hbox{Im}(W_{n})=0$. We list these 
points in order of decreasing phase (see Figure 5 and 6): 
\begin{eqnarray*} 
k &=&\left\{ 4q+1,2q,4q,2q-1,4q-1,....,q+1,3q+1\right\} \qquad \qquad\qquad
\hbox{for}\quad n=4q+1 \\ 
k &=&\left\{4q+3,2q+1,4q+2,2q,4q+1,2q-1,....,3q+3,q+1\right\} 
\quad \hbox{for}\quad n=4q+3 
\end{eqnarray*} 
The list of points $W_{h}$ for which $\hbox{Im}(W_{h})<0$ in order of 
decreasing phase is: 
\begin{eqnarray*} 
h &=&\left\{ q,3q,q-1,3q-1,.....,1,2q+1\right\} \qquad \qquad\qquad\quad
\hbox{for}\quad n=4q+1 \\ 
h &=&\left\{ 3q+2,q,3q+1,q-1,3q,.....,1,2q+2\right\} \qquad\qquad
\hbox{for}\quad n=4q+3 
\end{eqnarray*} 
All $W_{k}$ are equally spaced, with an angle of $2\pi /n$ between any 
two consecutive vacua. Let us look first at rays connecting vacuum 
$W_{0}$ with any other vacuum.  As $t$ is rotated, consecutive rays 
alternate between those of the form $\langle 0,k\rangle$ and those of 
the form $\langle h,0\rangle$. 
For $n=4q+1$ the sequence of rays passing through $W_{0}$ 
has the following form, ordered as 
they appear in the computation of $S$: 
\[\dots \langle 0,4q+1-j\rangle, \langle q-j,0 \rangle, 
\langle 0,2q-j\rangle,\langle 3q-j,0\rangle\dots\] 
The $A$ matrices corresponding to consecutive rays can 
be related by: 
\begin{eqnarray*} 
-A_{0,4q+1-j}^{T}&=&R^{-q}A_{q -j,0}R^{q}\\ 
-A_{q -j,0}^{T}&=&R^{-q}A_{0,2q -j}R^{q}\\ 
-A_{0,2q-j}^{T}&=&R^{-q}A_{3q -j,0}R^{q} 
\end{eqnarray*} 
and so on. 
For $n=4q+3$ the sequence is: 
\[\dots \langle 0,4q+3-j\rangle,\langle 3q+2-j,0\rangle, 
\langle 0,2q+1-j\rangle,\langle q-j,0 \rangle,...\] 
and the corresponding relations are: 
\begin{eqnarray*} 
-A_{0,4q+3-j}^{T}&=&R^{q+1}A_{3q+2 -j,0}R^{-q-1} 
\\ 
-A_{3q +2 -j,0}^{T}&=&R^{q+1}A_{0,2q+1 -j}R^{-q-1} 
\\ 
-A_{0,2q+1 -j}^{T}&=&R^{q+1}A_{q -j,0}R^{-q-1} 
\end{eqnarray*} 
The same connection exists between 
the matrices $A$ corresponding to lines which do not pass through 
vacuum $W_{0}$. We show this for $n=4q+1$. 
Any point $k$ for which 
$\hbox{Im}(W_{k})>0$ is of the form $k=2q -j$ or $k=4q+1-j$, while 
for a point $W_{h}$ for which $\hbox{Im}(W_{h})<0$, $h$ can be of the form $h=q -l$ 
or $h=3q -l$. Consider rays $\langle 2q-j,q-l \rangle$ 
and $\langle 2q-l, q-j \rangle$. These rays 
are parallel, as is easy to check by substituting 
the explicit expression for $W_k$ and $W_h$. 
Their $A$ matrices therefore both contribute to the same set 
$\mu ^{<s>}$. 
The lines contributing to the next set $\mu ^{<s+1>}$ then includes 
$\langle 3q-l,2q -j \rangle$ and all lines 
parallel to it. Each $A$ matrix in $\mu^{(s)}$ is 
related to one in $\mu^{(s+1)}$ by a relation like 
\[ 
-A_{q -j,2q -l}^{T}=R^{-q}A_{3q -l,2q -j}R^{q} .
\] 
Therefore the relation between a set $\mu ^{<s+1>}$ and $\mu 
^{<s>}$ is: 
\[ 
(I+\mu ^{<s>})^{T}=R^{-q}(I-\mu ^{<s+1>})R^{q} .
\] 
Using this relation, the 
matrix $S$ becomes 
\[ 
S=\left[ (I-\mu ^{<1>})R^{q}(I+\mu ^{<1>})^{T}R^{q}\right] ^{2q}(I-\mu 
^{<1>})R^{q}, 
\] 
where $\mu ^{<1>}$ is the sum of all $A$ matrices 
which appear first when computing $S$. These correspond to 
the parallel lines with 
the closest slope to $\pi $, along with the ray passing through the vacua 
$W_{0}$ and $W_{n}$. Because $(I-\mu ^{<1>})^{-T}=(I+\mu 
^{<1>})^{T}$, the monodromy matrix reduces to the form given in 
equation (\ref{s1}). For $n=4q+3$ the manipulations are similar and the 
monodromy matrix $M$ is given in equation (\ref{s2}).

We can finally solve for the soliton numbers in (\ref{detM}) by 
using the explicit expressions for the monodromy matrix coming 
from (\ref{s1}) or (\ref{s2}).  The Ramond charges (\ref{ramond})
are symmetrically distributed between $-3{c}/2$ and $3{c}/2$ 
(where $c=3n/(n+2)$ is the central charge of the conformal theory). 
Thus the $\lambda_{m}=\exp(2\pi 
iq_{m}^{R})$ and hence the eigenvalues of $M$ 
are all phases and distinct. This means that the eigenvalues 
$\beta_{m}$ of $M^{1/n}\equiv
(I-\mu ^{<1>})R^{q}(I+\mu ^{<1>})^{T}R^{q}$ are 
phases and distinct as well. This means that $M^{1/n}$ is also a 
monodromy, that is, it does not depend on perturbations. In fact,
both the group of automorphisms $Aut\mathtt{\ }H_{1}(V_{t})$ of 
the homology group $H_{1}(V_{t})$ and the fundamental group 
$\pi_{1}(t)$ factorize under $\mathbb{Z}_{n}$ into $n$ identical 
subgroups. Then the image of the homomorphism of such a subgroup 
of $\pi_{1}(t)$ in a subgroup of $Aut\mathtt{\ }H_{1}(V_{t})$ 
defines a monodromy. When all critical values are identical the 
eigenvalues of $(I-\mu ^{<1>})R^{q}(I+\mu ^{<1>})^{T}R^{q}$ are 
again given as exponentials of the Ramond charges.
This is consistent with the fact that the eigenvalues of $M$
are the same exponentials of Ramond charges, because
the set $(\beta_m)^n$ is the same as the
set of $\lambda_m$.
Therefore the eigenvalues of $M$ and $M^{1/n}$ are identical,
and we can substitute the latter for the former when solving (\ref{detM}). 
We can transform 
$\det(z-M^{1/n})$ by using the 
facts that $\det (I-\mu ^{<1>})=\det(R)=1$:
\begin{eqnarray*} 
\det\left[z- (I-\mu ^{<1>})R^{q}(I+\mu ^{<1>})^{T}R^{q}\right] 
&=&\det\left[ z R^{-q}(I-\mu ^{<1>})^{T}-(I-\mu 
^{<1>})R^{q}\right]\\ 
&=&\det\left\{R^{q}\left[ z R^{-q}(I-\mu ^{<1>})^{T}-(I-\mu 
^{<1>})R^{q}\right]R^{-q}\right\}\\ 
&=&\det\left[ 
z\left[ R^{q}(I-\mu ^{<1>})\right] ^{T}-R^{q}(I-\mu ^{<1>})\right] .
\end{eqnarray*} 
After doing the same transformation for $n=4q+3$, the equations to 
be solved for soliton numbers become: 
\begin{eqnarray*} 
\det\left[ z\left[ R^{q}(I-\mu ^{<1>})\right] ^{T}-R^{q}(I-\mu ^{<1>})\right] 
&=&P(z),\quad \hbox{ for }\quad n=4q+1 \\ 
\det\left[ z\left[ R^{-q-1}(I-\mu ^{<1>})\right] ^{T}-R^{-q-1}(I-\mu 
^{<1>})\right] &=&P(z),\quad \hbox{ for }\quad n=4q+3 .
\end{eqnarray*}

We can now use the explicit expressions for $\mu ^{<1>}$. For $n=4q+1$,
it is 
\begin{eqnarray*} 
(\mu ^{<1>})_{ij} &=&\delta _{i,0}\delta _{j,4q+1}\mu (0)-\delta 
_{i,q}\delta _{j,4q+1}\mu (q)-\delta _{i,3q+1}\delta _{j,2q}\mu 
(q+1)-\delta _{i,q+1}\delta _{j,4q}\mu (q+2)\\ 
&&-\delta _{i,3q+2}\delta_{j,2q-1}\mu (q+3)-\delta _{i,q+2}\delta 
_{j,4q-1}\mu (q+4)-\dots-\delta _{i,3q}\delta _{j,2q+1}\mu 
(q-1)\\ 
&&-\delta _{i,q-1}\delta_{j,1}\mu(q-2)-\delta _{i,3q-1}\delta 
_{j,2q+2}\mu (q-3)-\delta _{i,q-2}\delta _{j,2}\mu (q-4)-\dots 
\end{eqnarray*} 
where we have used $\mu (-i)=-\mu (i)$ 
and $\mu (i-n)=\mu (i)$. The matrix is of the form 
\[ 
I-\mu ^{<1>}=\left( 
\begin{array}{llllll} 
1 &  & & &  & -\mu (0) \\ 
& B & & &  &\ A \\ 
& & I_{q} &  & C & \\ 
& &  & D &  & \\ 
&  & E & & I_{q} & \\ 
&  & &  &  &\ 1 
\end{array} 
\right) , 
\] 
where omitted entries are zero, $I_{q}$ is the $q\times q$ identity, 
$A$ is a column of length $q$ and $B,C,D,E$ are $ 
q\times q$ matrices. For example for $q=5$, $n=21$ they are: 
\[ 
A=\left( 
\begin{array}{l} 
0 \\ 
0 \\ 
0 \\ 
0 \\ 
\mu (5) 
\end{array} 
\right) ,B=\left( 
\begin{array}{lllll} 
1 & 0 & 0 & 0 & 0 \\ 
0 & 1 & 0 & 0 & 0 \\ 
0 & \mu (1) & 1 & 0 & 0 \\ 
\mu (3) & 0 & 0 & 1 & 0 \\ 
0 & 0 & 0 & 0 & 1 
\end{array} 
\right) ,C=\left( 
\begin{array}{lllll} 
0 & 0 & 0 & 0 & \mu (7) \\ 
0 & 0 & 0 & \mu (9) & 0 \\ 
0 & 0 & 0 & 0 & 0 \\ 
0 & 0 & 0 & 0 & 0 \\ 
0 & 0 & 0 & 0 & 0 
\end{array} 
\right) 
\] 
\[ 
D=\left( 
\begin{array}{lllll} 
1 & 0 & 0 & 0 & 0 \\ 
0 & 1 & 0 & 0 & 0 \\ 
0 & 0 & 1 & 0 & 0 \\ 
0 & \mu (2) & 0 & 1 & 0 \\ 
\mu (4) & 0 & 0 & 0 & 1 
\end{array} 
\right) ,E=\left( 
\begin{array}{lllll} 
0 & 0 & 0 & 0 & \mu (6) \\ 
0 & 0 & 0 & \mu (8) & 0 \\ 
0 & 0 & \mu (10) & 0 & 0 \\ 
0 & 0 & 0 & 0 & 0 \\ 
0 & 0 & 0 & 0 & 0 
\end{array} 
\right) 
\] 
Multiplication to the left by $R^{q}$ shifts the rows, except 
for row $0$. The computation of the 
determinant gives: 
\begin{eqnarray*} 
P(z) &=&z^{n+1}-z^{n}(1-\mu (2q))-z(1-\mu 
(2q))+1-z^{\frac{n+1}{2}}(-\mu^{2}(0)-2\mu(q))+\\ 
& 
&+\sum_{k=1}^{2q}z^{n-2k+1}(-\mu(k)-\mu(2q-k+1))-\sum_{k=1,\mathtt{\ 
} k\neq q}^{2q-1}(-\mu(k)-\mu(2q-k)) .
\end{eqnarray*} 
By comparing the above polynomial with the form of $P(z)$ given by 
(\ref{detM}), one finds
\begin{eqnarray}
\nonumber
\mu (0)&=&\pm 1 \\ 
\label{solnum}
\mu (1)&=&\mu (2)=\mu (3)=....=\mu (q)=-1 \\
\nonumber
\mu (q+1)&=&\mu (q+2)=....=\mu (2q+1)=0 
\end{eqnarray}

The computation for $n=4q+3$ is similar. One has 
\begin{eqnarray*} 
(\mu ^{<1>})_{ij} &=&\delta _{i,0}\delta _{j,4q+3}\mu (0)+\delta 
_{i,q+1}\delta _{j,2q+1}\mu (q)+\delta _{i,3q+2}\delta 
_{j,4q+3}\mu (q+1)+\delta _{i,3q+3}\delta _{j,4q+2}\mu 
(q-1)+\\ 
&&+\delta _{i,q+2}\delta_{j,2q}\mu (q-2)+\delta _{i,3q+4}\delta 
_{j,4q+1}\mu (q-3)+\delta _{i,q+3}\delta _{j,2q-1}\mu 
(q-4)+\dots+\\ 
&&+\delta _{i,q}\delta _{j,2q+2}\mu (q+2)+\delta _{i,3q+1}\delta 
_{j,1}\mu(q+3)+\delta _{i,q-1}\delta _{j,2q+3}\mu (q+4)+\dots
\end{eqnarray*} 
Then $I-\mu ^{<1>}$ has the form: 
\[ 
I-\mu ^{<1>}=\left( 
\begin{array}{llll} 
I_{q+1} &  & A & B \\ 
& C &  & \\ 
D & & I_{q+1} & F \\ 
 & & & E 
\end{array} 
\right) , 
\] 
where $A,B,C,D,E,F$ are $(q+1)\times (q+1)$ 
matrices. Taking the determinant yields 
\begin{eqnarray*} 
P(z) 
&=&z^{n+1}-z^{n}(1-\mu (2q+1))-z(1-\mu 
(2q+1))+1+z^{\frac{n+1}{2}}(\mu^{2}(0)-2\mu(q+1))+\\ 
& &+\sum_{k=1,\mathtt{\ } k\neq 
q+1}^{2q+1}z^{n-2k+1}(-\mu(k)-\mu(2q-k+2))-\sum_{k=1}^{2q}z^{n-2k}(-\mu(k)-\mu(2q-k+1)) .
\end{eqnarray*} 
This yields soliton numbers identical to those for $n=4q+1$, as given
in (\ref{solnum}).

\begin{figure}[ht] 
\centerline{\includegraphics[scale=0.3,height=6cm]{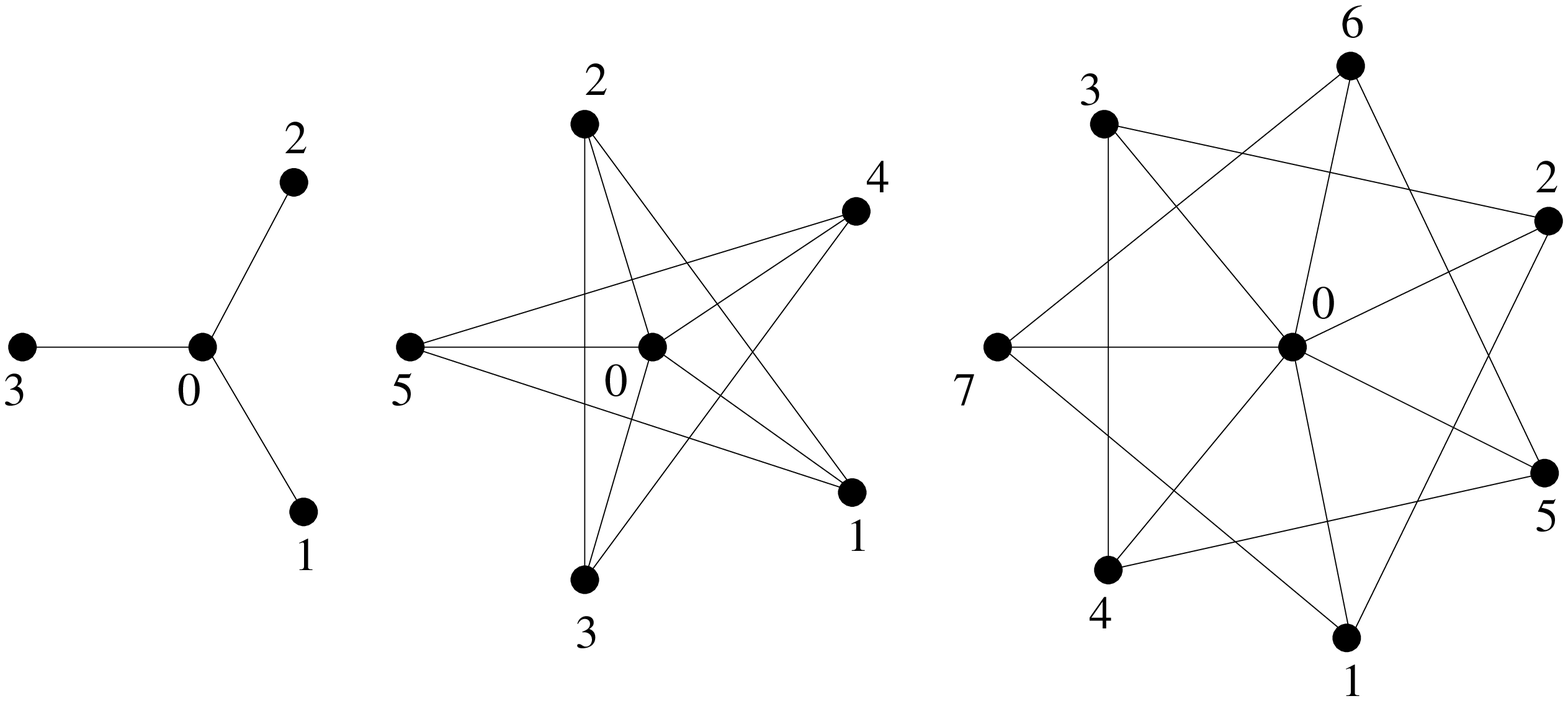}} 
\caption{Solitons in $W$-plane for $n=3,5,7$. Straight lines 
represent BPS states.} 
\end{figure} 
 
We have finally arrived at our answer for $n$ odd. For the first three 
cases, the soliton numbers are displayed graphically in figure 7; we 
draw a solid line to denote each doublet.  The pattern is fairly 
obvious. There is always a single doublet of fractional charges 
$(1/2,-1/2)$ connecting the origin to each of the other vacua.  For 
solitons not involving the origin, there is a single doublet 
connecting each vacuum to its $2[n/4]$ nearest neighbors {\it in the 
$X$ plane}. Nearest neighbors in the $X$ plane are not necessarily 
nearest neighbors in the $W$ plane, as is evident from the 
figure. Note also that the soliton doublets not involving the vacuum 
$0$ have either fractional charge ($0,1$) or ($0,-1$). $CP$ invariance 
means that if the doublet $(ab)$ has charge $(0,1)$, then the doublet 
$(ba)$ must have charges $(0,-1)$. The $\mathbb{Z}_n$ symmetry means 
that all clockwise solitons (say) are the former, all counterclockwise 
solitons are the latter. 
 
\section{The BPS spectrum for $n$ even} 
 
Finding the BPS spectrum for $n$ even requires more results from 
Picard-Lefschetz theory, because the critical values in $W$-plane 
are not all distinct. Two critical points diametrically opposed in 
the $X$ plane map into the same critical value in $W$ plane. 
Therefore two vanishing cycles $\Delta _{i},\Delta _{i+n/2}$ 
correspond to each $W_{i}$ with $i=1\dots n/2$, while one 
vanishing cycle $\Delta _{0}$ corresponds to the critical value at 
the origin in $W$-plane. The intersection matrix of two vanishing 
cycles corresponding to two critical points that map into the same 
critical value is zero, so $\Delta_i\circ\Delta_{i+n/2}= \mu 
\left( \left[ \frac{n}{2}\right] \right) =0$. This can also be 
seen directly from equation (\ref{solitoneqn}), because a BPS 
soliton satisfying $\Delta W=0$ must in fact be a constant value 
of $X$.

For $n$ even, the homology group splits naturally into two subspaces.  As 
before, the homology group $H_{1}(V_{t})$ of the non-singular level 
manifold $V_{t}\equiv W^{-1}(t)$ consists of $n+1$ vanishing cycles 
$\left\{ \Delta _{0},\Delta _{1},...,\Delta _{n}\right\} $, 
corresponding to the $n+1$ critical points.  However, here the 
superpotential $W$ is invariant under the involution $\sigma$ on $ 
\mathbb{C}^{2}$: $\sigma (X,Y)=(-X,Y)$.  The involution $\sigma $ acts 
on the vanishing cycles as \cite{picard2}: 
\begin{equation} 
\sigma _{*}\Delta _{i}=\Delta _{i+n/2},\quad \sigma _{*}\Delta _{0}=-\Delta 
_{0} 
\label{invo} 
\end{equation} 
This shows that the cycles $\Delta_{i}$ and $\Delta_{i+n/2}$ have different 
orientations.  Since $ \mathbb{Z}_{2}$ has two real 
irreducible representations, the homology group $H_{1}(V_{t})$ 
decomposes into a direct sum $H^{+}\oplus H^{-}$, where $ H^{+} $ 
contains the vanishing cycles invariant under the involution $\sigma 
$, and $H^{-}$ the anti-invariant ones. 
A basis in $H^{+}$ is given by `long-cycles' $\widetilde{ 
\Delta }_{i}$ with $i=1\dots n/2$: 
\[ 
\widetilde{\Delta }_{i}\equiv\Delta _{i}+\Delta _{i+n/2} 
\] 
while a basis in $H^{-}$ is given by `short-cycle' $\Delta _{0},$ 
and long-cycles $\widehat{\Delta }_{i}$ with $i=1\dots n/2$: 
\[ 
\widehat{\Delta }_{i}\equiv\Delta _{i}-\Delta _{i+n/2} 
\] 
Under the $\mathbb{Z}_{2}$, these cycles obey $\sigma 
_{*}\widetilde{\Delta }_{i}=\widetilde{\Delta }_{i}$ and $\sigma 
_{*}\widehat{\Delta }_{i}=-\widehat{\Delta }_{i}$, respectively. 
Intersections between cycles corresponding to different subgroups 
of the homology group are zero: 
\begin{eqnarray*} 
\Delta _{0}\circ \widetilde{\Delta }_{i} &=&0 \\ 
\widehat{\Delta }_{i}\circ \widetilde{\Delta }_{j} &=&0 
\end{eqnarray*} 
The change of basis formula in Picard-Lefschetz theory depends on
whether the cycle being crossed is long or short \cite{picard2,arnold}.  
When the path $\gamma_{a}$ to a vacuum $W_{a}$ crosses a long cycle
$\gamma_{b}$, $b\ne 0$, the change of basis $\alpha_{b}(a)$ is
\[ 
\Delta_{a}\rightarrow\Delta_{a}^{\prime }=\Delta_{a}-(\Delta_{a}\circ 
\Delta _{b})\Delta _{b}/2, 
\] 
whereas when it crosses a short cycle, the change of basis $\alpha_{0}(a)$ 
is 
\[ 
\Delta_{a}\rightarrow\Delta_{a}^{\prime }=\Delta_{a}-(\Delta_{a}\circ 
\Delta_{0} )\Delta_{0} , 
\]

One can associate the invariant and anti-invariant cycles to 
critical points of different functions. Let us consider the 
transformation: $\widehat{X} =X^{2},\widehat{Y}=Y$. Then 
$\widehat{X},\widehat{Y}$ are coordinates on the manifold 
$\mathbb{C}^{2}/\mathbb{Z}_{2}\simeq \mathbb{C}^{2}$ with boundary 
at $ \widehat{X}=0$. The two-dimensional hyperplane 
$\mathbb{C}^{2}$ with coordinates $X,Y$ can be viewed as the 
double covering of $\mathbb{C} ^{2}/\mathbb{Z}_{2}\simeq 
\mathbb{C}^{2}$ with a branch on the boundary $\widehat{X} =0.$ 
This transformation induces the function on 
$\mathbb{C}^{2}/\mathbb{Z}_{2}$: 
\[ 
\widehat{W}(\widehat{X}^{2},\widehat{Y})\equiv W(X,Y),\mathtt{\ 
}\hbox{ with }\mathtt{\ } 
\widehat{W}(\widehat{X},\widehat{Y})\equiv \frac{\widehat{X}^{(n+2)/2}}{n+2} 
-\lambda \frac{\widehat{X}}{2}+\widehat{Y}^{2} 
\] 
It can be shown \cite{picard2,wall} that the number of critical points (the 
Milnor multiplicity) $\nu $ of functions $W$ and $\widehat{W}$ satisfy: 
\[ 
\nu (W)=2\nu (\widehat{W})+\nu \left( \widehat{W}\mid \left\{ \widehat{X} 
=0\right\} \right) , 
\] 
and: 
\[ 
\dim H^{+}=\nu (\widehat{W})=n/2, 
\] 
\[ 
\dim H^{-}=\nu (\widehat{W})+\nu \left( \widehat{W}\mid \left\{ \widehat{X} 
=0\right\} \right) =n/2+1 
\] 
It is known that the analysis of singularities of functions ($W$) invariant 
under $\mathbb{Z}_{2}$ is equivalent to the analysis of singularities of 
function ($\widehat{W})$ on manifolds with boundary \cite{picard1,picard2}. 
The analysis of singularities of boundary singularity, that is of the 
function $\widehat{W}+\widehat{W}\mid _{\widehat{X}=0},$ can be carried out 
with the anti-invariant vanishing cycles, while the invariant cycles can be 
used for the singularities of $\widehat{W}.$ 
The characteristic polynomial $ 
P(z)$ factorizes into two polynomials $P^{+}(z)$ and $P^{-}(z)$ 
corresponding to the invariant and anti-invariant part respectively: 
\[ 
P(z)=P^{+}(z)P^{-}(z) 
\] 
Intersection numbers of cycles in different subgroups vanish, so 
the monodromy matrix $M$ can be computed 
independently on each subspace of the homology group of vanishing cycles. It 
has the form: 
\[ 
M=\left( 
\begin{array}{ll} 
M^{+} & 0 \\ 
0 & M^{-} 
\end{array} 
\right), 
\] 
where $M^{+}$ is an $n/2 \times n/2$ matrix and $M^{-}$ an $ 
(n/2+1)\times(n/2+1)$ matrix. 
Then, equation (\ref{detM}) becomes: 
\begin{eqnarray*} 
\label{detM-} 
\det(M^{-}-z)&=&P^{-}(z)\\ 
\det(M^{+}-z)&=&P^{+}(z) 
\label{detM+} 
\end{eqnarray*} 
where the polynomials are 
\begin{equation} 
P^{+}(z)=z^{n/2}-z^{n/2-1}+z^{n/2-2}-\dots \mp 1,\qquad\quad 
P^{-}(z)=z^{n/2+1}\pm1 \label{poly} 
\end{equation} 
with the last signs $(-,+)$ for $n=4q+2$ and $(+,-)$ for $n=4q$. 
To write the monodromy matrices $M^{+},M^{-}$ in terms of the 
soliton numbers, we need to look at vacua in $ W$-plane. There is 
a difference in the geometry of vacua in $W$-plane in the cases 
$n=4q+2$ and $n=4q$. When $n=4q$ there are three different vacua 
which are collinear, situation that does not appear in the case 
$n=4q+2$. Due to this difference we treat the two cases 
separately. 
 
\subsection{BPS spectrum for $\ n=4q+2$} 
 
The homology group splits in two, and the intersection numbers for 
cycles in different subspaces is zero. Finding the soliton numbers 
for $n=4q+2$ splits into two problems, one of which in fact has 
been solved before.

One additional complication is that the soliton numbers are of 
course invariant under the involution $\sigma$, but as detailed 
above, the intersection numbers are not (because the vanishing 
cycles are not (\ref{invo})). This requires a slight modification 
of the relation between the two. The trajectory between vacua 
$X_{0}$ and $X_{i}$ with $1\leq i\leq n/2$, and that between 
$X_{0}$ and $X_{i+n/2}$ map into the same straight line in 
$W$-plane. Since $\Delta _{i}$ and $\Delta _{i+n/2}$ have 
different orientations, $\Delta _{0}\circ \Delta _{i}$ and $\Delta 
_{0}\circ \Delta _{i+n/2}$ count the intersections points with 
different signs. Thus 
$$ 
\mu (0)\equiv \mu _{0i}=\mu_{0,i+n/2}= \Delta _{0}\circ \Delta _{i} 
= -\Delta _{0}\circ \Delta _{i+n/2}. 
$$ 
The situation is similar for the other solitons. When defining the 
soliton number between two critical points $W_{i}$ and $W_{k}$, 
with $1\le i \le n/2$, and $n/2<k\le n$, one must again include a 
minus sign. This yields 
\begin{eqnarray*} 
\mu _{ij}&=&\Delta _{i}\circ \Delta _{j},\quad \hbox{ for }\quad 1\leq i,j\leq n/2\\ 
\mu _{ik}&=& 
-\Delta _{i}\circ \Delta _{k},\quad \hbox{ for }\quad 1\leq i\leq n/2, \quad n/2<k\leq n,\\ 
\mu _{kl}&=&\Delta _{k}\circ \Delta _{l},\quad \hbox{ for }\quad n/2<k,l\leq n 
\end{eqnarray*} 
 
The model has a $\mathbb{Z}_n$ symmetry as before, although 
in the $W$ plane this is only $\mathbb{Z}_{n/2}$. 
These symmetries relate the soliton numbers $\mu_{ij}$, as before. 
On the invariant space,
\begin{eqnarray*} 
\Delta _{0}\circ \widetilde{\Delta }_{i}&=&0\\ 
\widetilde{\Delta }_{i}\circ \widetilde{\Delta }_{j}&=&2\mu (j-i)-2\mu 
(j-i-n/2) 
\end{eqnarray*} 
which yields 
\begin{eqnarray*} 
\widetilde{\Delta }_{i}\circ \widetilde{\Delta }_{i+p}&=&2\mu (p)+2\mu 
(2q-(p-1))\equiv 2\widetilde{\mu }(p)\\ 
\widetilde{\Delta }_{2q+1}\circ \widetilde{\Delta }_{1}&=&-2\mu (1)-2\mu (2q) 
\end{eqnarray*} 
for $p=1\dots q$ such that $1\leq i+p\leq 2q+1$. 
Therefore in the invariant part there are $q$ `soliton numbers' $\widetilde{ 
\mu }$,\ $\widetilde{\mu }_{ij}\equiv \frac{1}{2}\widetilde{\Delta } 
_{i}\circ \widetilde{\Delta }_{j}$ which satisfy: 
\begin{eqnarray*} 
\widetilde{\mu }_{ij}&=&\widetilde{\mu }(j-i)\\ 
\widetilde{\mu }(k)&=&-\widetilde{\mu }\left( k-\frac{n}{2}\right) 
\end{eqnarray*} 
for $k=1\dots q$. 
On the anti-invariant space the intersection numbers are: 
\begin{eqnarray*} 
\Delta _{0}\circ \widehat{\Delta }_{i}&=&2\mu (0)\\ 
\widehat{\Delta }_{i}\circ \widehat{\Delta }_{j}&=&2\mu (j-i)+2\mu 
(j-i-n/2) 
\end{eqnarray*} 
which give: 
\begin{eqnarray*} 
\widehat{\Delta }_{i}\circ \widehat{\Delta }_{i+p}&=&2\mu (p)-2\mu 
(2q-(p-1))\equiv 2\widehat{\mu }(p)\\ 
\widehat{\Delta }_{2q+1}\circ \widehat{\Delta }_{1}&=&2\mu (1)-2\mu (2q) 
\end{eqnarray*} 
for all $p=1\dots q$ such that $1\leq i+p\leq 2q+1$. 
Therefore on the anti-invariant part there are $q+1$ 'soliton 
numbers': $\mu (0)$ and  $\widehat{\mu }$, 
with $\widehat{\mu }_{ij}\equiv \frac{1}{2}\widehat{\Delta }_{i}\circ \widehat{ 
\Delta }_{j}$. These  satisfy: 
\begin{eqnarray*} 
\widehat{\mu }_{ij}&=&\widehat{\mu }(j-i)\\ 
\widehat{\mu }(k)&=&\widehat{\mu }\left( k-\frac{n}{2}\right) ,\mathtt{\ } 
\end{eqnarray*} 
for $k=1\dots q$.

The above properties of the $\widetilde{\mu }^{\prime }s$ are exactly the same 
as those satisfied by the real soliton numbers corresponding to the 
superpotential $\widehat{W}(\widehat{X},\widehat{Y})=\frac{\widehat{X} 
^{(n+2)/2}}{n+2}-\lambda \frac{\widehat{X}}{2}+\widehat{Y}^{2}$. These 
soliton numbers are known from the exact solution of this model in 
\cite{FI2}, and have been computed in \cite{CV2} using the Picard-Lefschetz 
techniques. In our conventions, the result is that all $\widetilde{\mu}=-1$. 
Thus the invariant part has been solved and, 
in terms of the original soliton numbers, we have: 
\begin{equation} 
\mu (p)+\mu (2q-(p-1))=-1,\mathtt{\ }p=1,...,q 
\end{equation}

\begin{figure}[ht] 
\centerline{\includegraphics[scale=0.4]{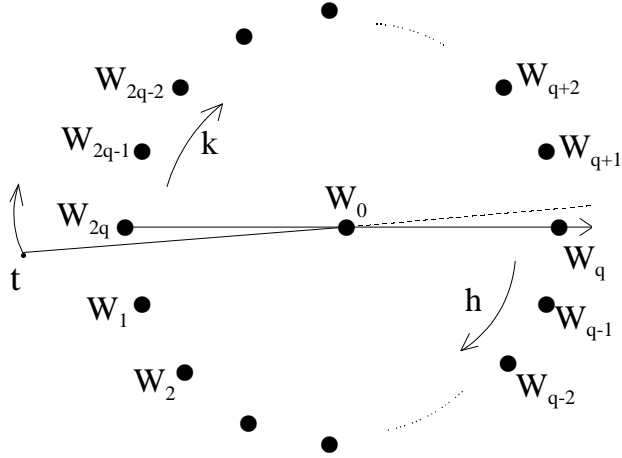}} \caption{Vacua 
geometry in $W$-plane for $n=4q+2$} 
\end{figure}

To compute the soliton numbers $\widehat{\mu }$ in the 
anti-invariant sector, we compute the monodromy matrix $M^{-}$ and 
use equation (\ref{detM-}). The vacua geometry in $W$-plane is 
represented in Figure 8.  We construct $M^-$ in the same way as we 
did for $n$ odd, but because of the factors of $2$ in various 
places, this will be slightly more complicated. Like before, the 
result can be simplified by relating basis matrices appearing 
sequentially.  Here the sequence of critical values in $W$-plane 
is 
\begin{eqnarray*} 
k &=&\left\{ 2q+1,2q,2q-1,2q-2,.....,q+3,q+2,q+1\right\} \\ 
h &=&\left\{ q,q-1,q-2,q-3,....,3,2,1\right\} 
\end{eqnarray*} 
where Im$(W_k)\ge 0$, while Im$(W_h)\le0$. The matrix $A_{0k}$ 
(which arises when $t$, $W_{0}$ and $W_{k}$ are collinear) has 
$\mu (0)$ for its only non-zero entry $(0,k)$. However the 
change-of-basis matrix arising when $t$, $W_{0}$ and $W_{h}$ are 
collinear contains the entry $2\mu (0)$. Let us therefore denote 
by $\widehat{A}_{h0}$ the matrix having $2\mu (0)$ at the entry 
$(h,0)$ and zero in rest. We can connect such $A$ and 
$\widehat{A}$ matrices, ordered as they appear when computing 
$M^{-}$, in the following way ($0\leq i\leq q-1 $): 
\begin{eqnarray*} 
R^{-q}\widehat{A}_{q-i,0}R^{q} &=&-2{A}_{0,2q+1-i}^{T} \\ 
R^{-q}{A}_{0,2q-i}R^{q} &=&-\frac{1}{2}\widehat{A}_{q-i,0}^{T} 
\end{eqnarray*} 
Let us consider now the lines connecting two vacua except the origin, 
so that this involves only long cycles.  The change of basis for when 
$t$, $W_{i}$ and $W_{j}$ are collinear is $M_{ij}=I-A_{ij}$, where 
$A_{ij}$ has $(i,j)$ the only non-zero entry equal to $\widehat{\mu 
}_{ij}$. The change of such basis matrices for such lines can be 
related again using matrices $R^{q}$ and $R^{-q}$. This yields 
the monodromy matrix: 
\[ 
M^{-}=\left[ \left( I-\mu _{1}^{\left\langle 1\right\rangle }\right) 
R^{q}\left( I+\mu _{2}^{\left\langle 1\right\rangle }\right) 
^{T}R^{q}\right] ^{n/2}, 
\] 
where: 
\[ 
\left( \mu _{1}^{\left\langle 1\right\rangle }\right) _{ij}=\delta 
_{i,0}\delta _{j,2q+1}\mu (0)+\left( \mu ^{\left\langle 1\right\rangle 
}\right) _{ij} 
\] 
\[ 
\left( \mu _{2}^{\left\langle 1\right\rangle }\right) _{ij}=2\delta 
_{i,0}\delta _{j,2q+1}\mu (0)+\left( \mu ^{\left\langle 1\right\rangle 
}\right) _{ij} 
\] 
As before, $\mu ^{<1>}$ is the sum of all matrices $A_{ij}$ 
corresponding to the parallel lines which appear first (after the ray 
connecting $W_{0}$ and $ W_{2q+1}$). Explicitly, it is 
\begin{eqnarray*} 
\left( \mu ^{\left\langle 1\right\rangle }\right) _{ij} &=&-\delta 
_{i,q}\delta _{j,2q+1}\widehat{\mu }(q)+\delta _{i,q+1}\delta _{j,2q} 
\widehat{\mu }(q-1)+\delta _{i,q+2}\delta _{j,2q-1}\widehat{\mu }(q-3)+ \\ 
&&+\delta _{i,q+3}\delta _{j,2q-2}\widehat{\mu }(q-5)+......-\delta 
_{i,q-1}\delta _{j,1}\widehat{\mu }(q-2)-\delta _{i,q-2}\delta _{j,2} 
\widehat{\mu }(q-4)- \\ 
&&-\delta _{i,q-3}\delta _{j,3}\widehat{\mu }(q-6)-\delta _{i,q-4}\delta 
_{j,4}\widehat{\mu }(q-8)-\dots
\end{eqnarray*} 
As for $n$ odd, solving equation (\ref{detM-}) for $\widehat{\mu }^{\prime 
}s$ reduces to solving: 
\[ 
\det\left[ \left( I-\mu _{1}^{\left\langle 1\right\rangle }\right) 
R^{q}\left( I+\mu _{2}^{\left\langle 1\right\rangle }\right) 
^{T}R^{q}-z\right] =P^{-}(z) .
\] 
Computing the determinant  yields 
\begin{eqnarray*} 
P^{-}(z) &=&z^{2q+2}+1-z^{2q+1}(1+\widehat{\mu }(1))-z(1+\widehat{\mu } 
(1))+z^{q+1}(2\mu ^{2}(0)+2\widehat{\mu }(q))+ \\ 
&&+\sum_{k=0}^{q-2}\left[ (-1)^{k}z^{2q-k}(\widehat{\mu }(k+1)-\widehat{\mu } 
(k+2))+(-1)^{q-k}z^{q-k}(\widehat{\mu }(q-k-1)-\widehat{\mu }(q-k))\right] .
\end{eqnarray*} 
Thus plugging in the polynomial 
from (\ref{poly}) 
\[ 
\mu (0)=\pm 1,\mathtt{\ }\widehat{\mu }(i)=-1,\mathtt{\ }i=1,\dots,q 
\] 
which in terms of the original $\mu ^{\prime }s$ give: 
\[ 
\mu (p)-\mu (2q-(p-1))=-1,\mathtt{\ }p=1,\dots,q .
\] 
Reexpressing this in terms of the original soliton numbers 
yields (after taking  absolute values) 
\[ 
\mu (0)=1 
\] 
\[ 
\mu (1)=\mu (2)=\mu (3)=....=\mu (\left[ \frac{n}{4}\right] )=1 
\] 
\begin{equation} 
\mu (\left[ \frac{n}{4}\right] +1)=\mu (\left[ \frac{n}{4}\right] 
+2)=....=\mu (\left[ \frac{n}{2}\right] )=0 
\end{equation} 
So this is essentially the same as for $n$ odd: there is a 
single BPS doublet joining the origin to each vacuum, and a single 
BPS doublet joining a vacuum to its $2q$ neighbors in the $X$-plane.

\subsection{BPS spectrum for $n=4q$}

The case $n=4q$ is the most complicated.  This is because the 
vacua $W_{0}$, $W_{i}$ and $W_{i+q}$ are collinear for any 
$i=1\dots q$. Collinear vacua result in contributions to the index 
$C$ from multi-particle states that get mixed in with those from 
one-particle states. Specifically, $\mu_{ab}$ is defined as the 
coefficient of $C_{ab}$ in the $\beta\to\infty$ limit, as 
shown in (\ref{IR}).  Since BPS solitons are constrained to have 
mass $m_{ab}=|\Delta_{ab}|$, the one-particle states indeed 
generically provide the only contributions to $C_{ab}$ at $\beta$ 
large. The one situation where the multi-particle contributions 
can be just as large is if there is some vacuum $c$ collinear with 
and in between $a$ and $b$. The central terms are additive in 
general: $\Delta_{ab}=\Delta_{ac} + \Delta_{cb}$ (this is obvious 
in a Landau-Ginzburg theory, where $\Delta_{ab}=W(b)-W(a)$). The 
special property of collinear vacua is that 
$|\Delta_{ab}|=|\Delta_{ac}| +|\Delta_{cb}|$.  Thus a two-particle 
BPS state $(ac)(cb)$ would have the same mass as a one-particle 
state $(ab)$, and can contribute to $C_{ab}$ in the same way. This 
means that $\mu_{ab}$ does not necessarily count the one-particle 
BPS doublets. In fact, $\mu_{ab}=1/2$ in the $q=1$ case, as was 
shown by explicit solution of the differential equation 
\cite{CV2}. There is no such thing as a half-soliton, so in this 
case this contribution must arise solely from the two-particle 
states. It is important to note that even though two-particle 
states are not necessarily in a BPS representation, they can 
contribute to $C_{ab}$ \cite{CFIV}; the argument in the 
introduction applies only to one-particle states.

For $n=4q$, the relations of the soliton numbers to the 
intersection numbers are essentially the same as in the case 
$n=4q+2$ given in the last subsection.  However, because of the 
collinearity, the definitions of $\mu _{i,i+q}$ and $\mu_{i,i-q}$ 
require special attention.  Let us analyze this situation in the 
context of singularity theory. Consider three such critical values 
$W_{0}$, $W_{i}$, $W_{i+n/4}$, with $t$ a non-critical point on 
the line between $W_{0}$ and $W_{i}$.  As before, $\gamma_i$ and 
$\gamma_0$ are straight lines connecting $t$ to $W_{i}$ and 
$W_{0}$, but with $\gamma_{i+n/4}$, we deform the line so that it 
makes a small semi-circle around the origin.  In $V_{t}$, the 
intersection between $\Delta _{0}$ and $\Delta _{i}$ gives $\mu 
_{0i}$. In addition, $\Delta _{i}$ and $\Delta _{i+n/4}$ 
intersect, but they may have common intersection points with 
$\Delta _{0}\circ \Delta _{i}$. The common intersection points 
give contributions which look like two-soliton states between 
vacua $X_{i}$ and $ X_{i+n/4}$ whose number we denote by $\mu 
_{i,i+n/4}^{2}=\mu^{2}(n/4)$, and the non-common intersection 
points give one-soliton states whose number is $\mu 
_{i,i+n/4}^{1}=\mu^{1}(n/4)$. The same analysis applies when 
considering the vacua $X_{i+n/4},X_{0}$ and $X_{i+n/2}$. The vacua 
$X_{i}$ and $X_{i+n/2}$ map into the same $W_{i}$, but to maintain 
the $\mathbb{Z}_{n}$-symmetry, the line between $W_{i}$ and 
$W_{i+n/4}$ avoids the origin by a small semi-circle opposite to 
the above semi-circle. We include a minus sign in the definition 
of $\mu ^{2}$ coming from the definitions of $\mu _{0i}$ and $\mu 
_{0,i+n/2}$, but not in the definition of $\mu ^{1}$ since the 
straight lines are not exactly the same. Therefore, when $\left| 
i-j\right| =|j-k|=q$, with $1\leq i,j\leq n/2$ and $n/2<k,l\leq 
n$, we define the soliton numbers as: 
\begin{eqnarray*} 
\Delta _{i}\circ \Delta _{j}=\mu _{ij}^{1}+\mu _{ij}^{2}\\ 
\Delta _{k}\circ \Delta _{l}=\mu _{kl}^{1}+\mu _{kl}^{2}\\ 
\Delta _{i}\circ \Delta _{k}=\mu _{ik}^{1}-\mu _{ik}^{2} 
\end{eqnarray*} 
This procedure of separating the intersection matrix into two 
components is equivalent to a partial diagonalization of the 
intersection matrix. The Dynkin diagram of a singularity of a 
function of two variables can be often represented in the form of 
the Dynkin diagram of a real curve. The connection between the two 
Dynkin diagrams is done by a partial diagonalization of the 
intersection matrix. It is known \cite{picard2} that the 
intersection numbers, after diagonalization, belong to the set of 
half-integer $\frac{1}{2}\mathbb{Z}$. We will indeed obtain 
intersection numbers of $1/2$ for three collinear vacua for which 
the above representation of Dynkin diagrams is valid. 
 
We proceed now, as in the $n=4q+2$ case, by expressing the 
intersections on invariant and anti-invariant spaces in terms of 
soliton numbers we need to compute. For the invariant space we 
have: 
\begin{eqnarray*} 
\widetilde{\Delta }_{i}\circ \widetilde{\Delta }_{i+p}&=&2\mu (p)+2\mu 
(2q-p)\equiv 2\widetilde{\mu }(p)\\ 
\widetilde{\Delta }_{i}\circ \widetilde{\Delta }_{i+q}&=&4\mu ^{2}(q)\equiv 2 
\widetilde{\mu}(q) 
\end{eqnarray*} 
where $p=1\dots q-1$. There are again $q$ different 
$\widetilde{\mu}(i)$ in the invariant part. As in the case 
$n=4q+2$, they are identified with the real soliton numbers for 
the superpotential 
$\widehat{W}(\widehat{X},\widehat{Y})=\frac{\widehat{X} 
^{(n+2)/2}}{n+2}-\lambda \frac{\widehat{X}}{2}+\widehat{Y}^{2}$. 
Since these are known to be all $-1$ \cite{FI2,CV2}, we have 
\begin{eqnarray} 
\label{invsol} 
\mu (p)+\mu (2q-p)&=&-1,\\ 
\mu ^{2}(q)&=&-1/2 
\label{halfsol} 
\end{eqnarray} 
The latter must be a two-soliton contribution, and we see how 
this contribution of $\pm 1/2$ to $\mu_{ab}$ arises naturally 
in singularity theory. We have thus proven directly that the behavior already 
observed for $q=1$ persists for all $q$. 
 
Finding the soliton numbers in the 
anti-invariant space requires a new computation. The soliton 
numbers are related to the intersection numbers by 
\[ 
\widehat{\Delta }_{i}\circ \widehat{\Delta }_{i+p}=2\mu (p)-2\mu 
(2q-p)\equiv 2\widehat{\mu }(p),\mathtt{\ }p=1,...,q-1 
\] 
\[ 
\widehat{\Delta }_{i}\circ \widehat{\Delta }_{i+q}=4\mu ^{1}(q)\equiv 2 
\widehat{\mu }(q) 
\] 
Note that the one-soliton 
number given by $\mu^{1}(q)$ is in the anti-invariant space. 
To compute $\mu (0)$ and $\widehat{\mu }$ we use again equation 
(\ref{detM-}). The vacua geometry in $W$-plane 
is represented in Figure 9, and the sequence of points is: 
\begin{eqnarray*} 
k &=&\left\{ 2q,2q-1,2q-2,.....,q+3,q+2,q+1\right\} \\ 
h &=&\left\{ q,q-1,q-2,....,3,2,1\right\} 
\end{eqnarray*} 
where again Im$(W(k))\ge 0$, and Im$(W(h))\le 0$. 
\begin{figure}[ht] 
\centerline{\includegraphics[scale=0.4]{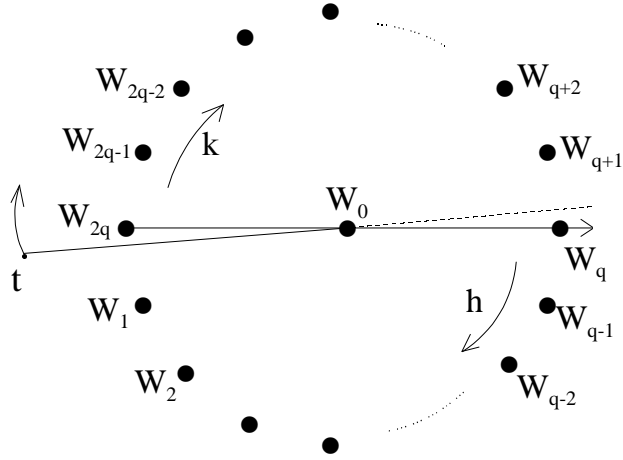}} \caption{Vacua 
geometry in $W$-plane for $n=4q$} 
\end{figure} 
Let us first study the change of basis matrices associated with 
the collinear vacua $W_{0},W_{q}$ and $W_{2q}$. As $t$ moves, the 
straight line paths $\gamma _{0},\gamma _{q}$ and $\gamma _{2q}$ 
will change into $\gamma _{0}^{^{\prime }},\gamma _{q}^{^{\prime 
}}$ and $\gamma _{2q}^{^{\prime }}$ as in Figure 10. 
It is known \cite{picard2,humph} that two weakly 
distinguished bases can be obtained one from the other by 
iterations of operations $\alpha _{a}(b)$ and $\beta _{a}(b)$. 
There are two such change of paths operations to change $\gamma $ 
into $\gamma ^{\prime }$: $\alpha _{0}(q)\alpha _{2q}(q)\alpha 
_{2q}(0)$ and $\alpha _{2q}(0)\alpha _{2q}(q)\alpha _{0}(q)$. 
However, they do not commute. 
\begin{figure}[ht] 
\centerline{\includegraphics[scale=0.4]{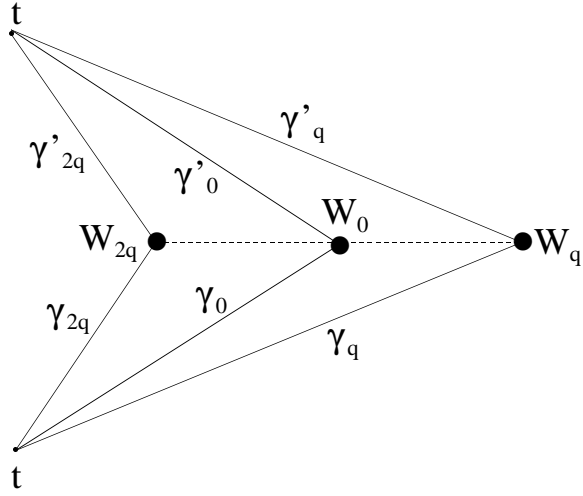}} 
\caption{Three collinear vacua in $W$-plane: $W_{q},W_{0},W_{2q}$} 
\end{figure} 
Since we want to obtain soliton numbers which do not depend on how 
the paths $\gamma $ transform into $\gamma ^{\prime }$, we take 
the change of basis matrix as being the symmetrized change of 
basis matrices of the two operations. For example, for the above 
change of paths the change of basis is given by: 
\[ 
M_{q,0,2q}\equiv\frac{1}{2}\left\{ 
(I-\widehat{A}_{0,2q}),(I-A_{q,0})\right\} (I-A_{q,2q}), 
\] 
where $A$ and $\widehat{A}$ have the same meaning as in the case $n=4q+2$, 
and the anti-commutator has the elements: 
\begin{eqnarray*} 
\frac{1}{2}\left\{ (I-\widehat{A}_{0,2q}),(I-A_{q,0})\right\} _{ij} &\equiv 
&\delta _{ij}-\left( \mu _{0}^{<1>}\right) _{ij}= \\ 
&=&\delta _{ij}+\delta _{i,q}\delta _{j,0}2\mu (0)-\delta _{i,0}\delta 
_{j,2q}\mu (0)-\delta _{i,q}\delta _{j,2q}\mu ^{2}(0) 
\end{eqnarray*} 
Such matrices $M$ can be connected in order as they appear in the following 
sense ($1\leq i\leq q-1$): 
\[ 
RM_{q-i,0,2q-i}R^{-1}=M_{q-i+1,0,2q-i+1} 
\] 
\[ 
RM_{2q,0,q}R^{-1}=-M_{1,0,q+1} 
\] 
\[ 
RM_{2q-i,0,q-i}R^{-1}=M_{2q-i+1,0,q-i+1} 
\] 
The sets of lines which do not pass through the origin can also be connected 
in the same way. For a given direction the parallel lines contain half of 
the soliton numbers but each one appears twice. However, we can connect such 
parallel lines by using $R^{q},R^{-q}$ to have each soliton number appearing 
once. Then, we obtain the monodromy matrix $M^{-}$: 
\[ 
M^{-}=-\left[ \left( I-\mu _{0}^{<1>}\right) \left( I-\mu 
_{2}^{<1>}\right) R^{q}\left( I+\mu _{1}^{<1>}\right)^{T} 
R^{q-1}\right] ^{n/2} 
\] 
where: 
\[ 
\left( \mu _{2}^{<1>}\right) _{ij}=\delta _{i,q}\delta _{j,2q}\widehat{\mu } 
(q)+\left( \mu _{1}^{<1>}\right) _{ij} 
\] 
and $\mu _{1}^{<1>}$ is the sum of all $A$ matrices which correspond to the 
first and second set of parallel lines (excluding $A_{q,2q}$) such that all 
soliton numbers appear once. It has the elements: 
\begin{eqnarray*} 
\left( \mu _{1}^{<1>}\right) _{ij} &=&\delta _{i,q}\delta _{j,2q-1}\widehat{ 
\mu }(q-1)+\delta _{i,q+1}\delta _{j,2q-2}\widehat{\mu }(q-3)+\delta 
_{i,q+2}\delta _{j,2q-3}\widehat{\mu }(q-5)+... \\ 
&&+\delta _{i,q+1}\delta _{j,2q-1}\widehat{\mu }(q-2)+\delta _{i,q+2}\delta 
_{j,2q-2}\widehat{\mu }(q-4)+\delta _{i,q+3}\delta _{j,2q-3}\widehat{\mu } 
(q-6)... 
\end{eqnarray*} 
Again, solving (\ref{detM-}) reduces to solving: 
\[ 
\det\left[ \left( I-\mu _{0}^{<1>}\right) \left( I-\mu _{2}^{<1>}\right) 
R^{q}\left( I+\mu _{1}^{<1>}\right) ^{T}R^{q-1}+z\right] =P^{-}(z) 
\] 
The computation of the determinant gives: 
\begin{eqnarray*} 
P^{-}(z) 
&=&z^{2q+1}-1-z^{2q}(-1-\widehat{\mu }(1))+z(-1-\widehat{\mu }(1))+ \\ 
&&+(-1)^{q}z^{q+1}(\mu ^{2}(0)+\widehat{\mu }(q-1)+\widehat{\mu } 
(q))+(-1)^{q-1}z^{q}(\mu ^{2}(0)+\widehat{\mu }(q-1)-\widehat{\mu }(q))+ \\ 
&&+\sum_{k=1}^{q-2}\left[ (-1)^{k+1}z^{2q-k}(\widehat{\mu }(k)-\widehat{\mu } 
(k+1))+(-1)^{q-k-1}z^{q-k}(\widehat{\mu }(q-k-1)-\widehat{\mu }(q-k))\right]. 
\end{eqnarray*}

Plugging in the polynomial 
from (\ref{poly}) gives us the solutions 
\begin{eqnarray*} 
\mu (0)&=&\pm 1,\\ 
\widehat{\mu }(i)&=&-1,\qquad\mathtt{\ }i=1,...,q-1\\ 
\widehat{\mu }(q)&=&0 
\end{eqnarray*} 
which gives the original soliton numbers as 
\begin{eqnarray*} 
\mu ^{1}(q)&=&0,\\ 
\mu (p)-\mu (2q-p)&=&-1,\quad\qquad p=1,...,q-1. 
\end{eqnarray*} 
Finally, combining this with the results for 
invariant cycles (\ref{invsol},\ref{halfsol}) yields 
\[ 
\mu (0)=1 
\] 
\[ 
\mu (1)=\mu (2)=\mu (3)=....=\mu (\left[ \frac{n}{4}\right] -1)=1,\mathtt{\ } 
\mu (\left[ \frac{n}{4}\right] )=1/2 
\] 
\begin{equation} 
\mu (\left[ \frac{n}{4}\right] +1)=\mu (\left[ \frac{n}{4}\right] 
+2)=....=\mu (\left[ \frac{n}{2}\right] )=0 
\end{equation} 
Note that there are no fundamental BPS doublets connecting vacua 
$i$ and $i+q$. because $\mu ^{1}(\left[ \frac{n}{4}\right] )$ is 
zero. As for $n$-odd, we display graphically the soliton numbers 
for several particular cases of $n$-even in figure 11. In $W$ plane the
symmetry is  only $\mathbb{Z}_{n/2}$. The pattern is clear: there is one line
(one  BPS doublet) connecting any two vacua except for the vacua which are
collinear  with the vacuum at the origin. 
\begin{figure}[ht] 
\centerline{\includegraphics[scale=0.3,height=6cm]{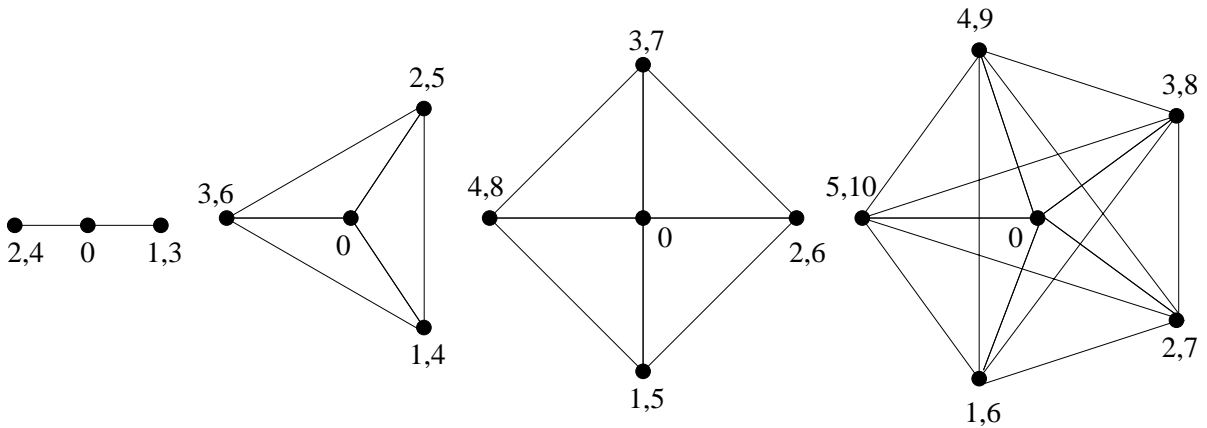}} 
\caption{Solitons in the $W$-plane for $n=4,6,8,10$. Straight lines 
represent BPS states.} 
\end{figure} 
 
We end the computation for $n$ even by noticing the similarity 
between this case and the most relevant perturbation of the $D$ 
series of minimal superconformal models
given by $W=X^{(n+2)/2}-XY^{2}-\lambda X$. In fact,
the two are related by an orbifold \cite{FLMW}. 
For this perturbed $D$ series one can do 
similar computations as in sections 5.1 and 5.2, and obtain 
exactly the same expressions for the determinants, and therefore 
the same soliton numbers. 
 
\section{Non-BPS states}

We have found the full BPS spectrum of the ${\cal N}=(2,2)$ 
Landau-Ginzburg theory $X^{n+2}-\lambda X^{2}$ for any $n$.  The 
Picard-Lefschetz theory of singularities of differentiable maps 
proved to be a powerful tool in finding the BPS spectrum exactly. 
This technique is more direct than using topological 
anti-topological fusion equations, because extracting the 
appropriate behavior from those equations explicitly is difficult 
at best.

There is one interesting consequence of this computation.  As we 
already mentioned, the theories studied here are integrable 
\cite{FLMW,MW}.  Integrable models have the striking property that 
the $ S$-matrix factorizes into two-body ones, and all scattering 
processes are completely elastic \cite{zamo1}. This means that in 
any scattering process, the masses of the {\it individual} 
particles are conserved. The only possible consequences of 
scattering are a phase shift or changing internal quantum numbers. 
This fact, along with the fact that the boundary conditions at 
spatial infinity remain fixed in a collision, means that we can 
show there must be non-BPS single-particle states. The argument 
for this is quite simple. Consider a two-particle configuration 
$(0a)(ab)$, where $a$ and $b$ are vacua other than the origin. The 
masses of these BPS particles are given by the length of the 
corresponding lines in the $W$ plane, namely $m_{0a}=|W(a)-W(0)|$, 
and $m_{ab}=|W(b)-W(a)|$. Such a two-particle configuration is 
drawn for $n=7$ in figure 12. When these two particles scatter, 
they must scatter to two particles $(0c)(cb)$. Since the theory is 
integrable, there can be no particle production, and we must 
have $m_{0c}=m_{ab}$ and $m_{cb}=m_{0a}$. Note that it is not 
possible for $c=a$; the solitons must change sides in the 
scattering event. In fact, in general there is no vacuum $c$ where 
$|W(c)-W(0)|=|W(b)-W(a)|$ or with $|W(b)-W(c)|=|W(a)-W(0)|$. This 
can be easily seen from the picture in figure 12: there is no line 
$(0c)$ of the same length as the line $(23)$, and no line $(c3)$ 
of the same length as $(02)$. This means that the final state in 
this scattering process cannot consist of BPS states! The theory 
must contain additional states, and the scattering of BPS states 
does not close onto itself.
 
\begin{figure} 
\centerline{\includegraphics[scale=0.39]{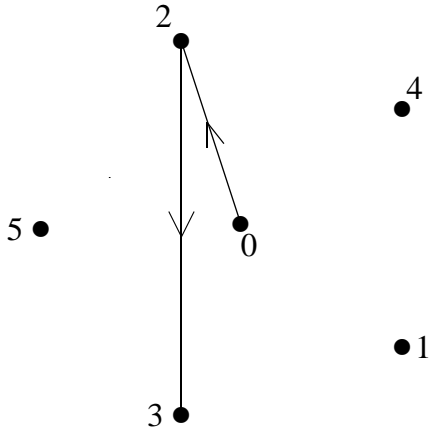}} \caption{
Two-soliton configuration in the $W$ plane when $n=5$} 
\end{figure} 

We have therefore proven that there are non-BPS states mixing
with the BPS states here, as opposed to the models discussed in
\cite{FI1,FI2}, where the entire spectrum consists only of BPS states. 
The simplest scenario consistent with integrability and the BPS spectrum 
is for there to exist non-BPS states $(00)$. Presumably, these 
states are analogous to the breathers well known in the sine-Gordon field 
theory. Because of the integrability, the masses of these non-BPS 
quartets must be the same as the BPS states, 
except a non-BPS particle of mass $m_{0i}$ is not required. 
We will give the $S$ matrix for all of these states in the sequel 
\cite{TF}. 
 
\bigskip\bigskip 
This work was supported in part by 
the Research Corporation under a Research Innovation Award, a DOE OJI award 
an NSF grant DMR-0104799, and a Sloan Foundation Fellowship.

\end{document}